%% file: Sum_Rule_New.tex
\documentclass[11pt]{article} 

\usepackage[utf8]{inputenc} 

\usepackage{geometry} 
\usepackage{graphicx} 

\usepackage{ bm }
\usepackage{ textcomp }
\usepackage{ amsmath }
\usepackage{ amssymb }
\usepackage{ bbm }
\usepackage{ authblk }

\usepackage[colorlinks=true, citecolor=blue]{hyperref} 
\usepackage[numbers,sort&compress]{natbib}
\usepackage{booktabs} 
\usepackage{array} 
\usepackage{paralist} 
\usepackage{verbatim} 
\usepackage{subfig} 

\usepackage{fancyhdr} 
\usepackage{sectsty}
\allsectionsfont{\mdseries\upshape} 

\usepackage[title]{appendix}

\usepackage[nottoc,notlof,notlot]{tocbibind} 
\usepackage[titles,subfigure]{tocloft} 

\title{Reciprocal space study of Heisenberg exchange interactions in ferromagnetic metals}
\author{I. V. Kashin$^{*}$}
\author{A. Gerasimov}
\author{V. V. Mazurenko}

\affil{\textit{Theoretical Physics and Applied Mathematics Department, Ural Federal University, Mira Str. 19, 620002 Ekaterinburg, Russia}}
\affil{$^{*}$Corresponding author: i.v.kashin@urfu.ru}

\begin{document}
\maketitle

\input{Regulars}

\input{Abstract}

\input{Introduction}

\input{Method}

\input{Exchange_Surrounding_Problem}

\input{Orbital_Decomposition_and_Symmetry_Problem}

\input{Conclusions}

\input{Acknowledgements}

\input{Appendix}

\input{Bibliography}

\end{document}

%% file: Regulars.tex
\newcommand{\regexp}{\mathrm{exp}}

\newcommand{\energy}{E}

\newcommand{\GkE}{{\cal{G}}^{\sigma}(\energy, \bm{k})}

\newcommand{\GkEup}  {{\cal{G}}^{\uparrow}  (\energy, \bm{k})}
\newcommand{\GkEdown}{{\cal{G}}^{\downarrow}(\energy, \bm{k})}
\newcommand{\GkprimeEdown}{{\cal{G}}^{\downarrow}(\energy, \bm{k'})}

\newcommand{\dE}{dE}

\newcommand{\ElecHamE}{{\cal{E}}}

\newcommand{\shortminus}{\text{-}}

\newcommand{\shorteq}{\text{\textdblhyphen \,}}

\newcommand{\sublati}{\tilde{i}}
\newcommand{\sublatj}{\tilde{j}}

\newcommand{\JqZero}{J(\bm{q} \, \shorteq 0)}

\newcommand{\bccFe}{\textit{bcc} Fe\,}

\newcommand{\ttwogttwog}{t_{2g} \shortminus t_{2g}}
\newcommand{\egeg}{e_g \shortminus e_g}
\newcommand{\ttwogeg}{t_{2g} \shortminus e_g}

\newcommand{\ParsP}{{\cal{P}}}
\newcommand{\ParsR}{{\cal{R}}}
\newcommand{\ParsQ}{{\cal{Q}}}

\newcommand{\CubicRootofNq}{\sqrt[\leftroot{-1}\uproot{2} 3]{N_{\bm{q}}}}

\newcommand{\QGrid}{\CubicRootofNq~\times~\CubicRootofNq~\times~\CubicRootofNq}

\newcommand{\frc}[2]{\raisebox{2pt}{$#1$}\big/\raisebox{-3pt}{$#2$}} 

\newenvironment{psmallmatrix}
  {\left(\begin{smallmatrix}}
  {\end{smallmatrix}\right)}

%% file: Abstract.tex
\noindent
The modern quantum theory of magnetism in solids is getting commonly derived using Green's functions formalism.
The popularity draws itself from remarkable opportunities to capture the microscopic landscape of exchange interactions, starting from a tight-binding representation of the electronic structure.
Indeed, the conventional method of infinitesimal spin rotations, considered in terms of local force theorem, opens vast prospects of investigations regarding the magnetic environment, as well as pairwise atomic couplings.
However, this theoretical concept practically does not devoid of intrinsic inconsistencies.
In particular, naturally expected correspondence between single and pairwise infinitesimal spin rotations is being numerically revealed to diverge.
In this work, we elaborate this question on the model example and canonical case of \textit{bcc} iron.
Our analytical derivations 
discovered the principal preference of on-site magnetic precursors 
if the compositions of individual atomic interactions are in focus.
The problem of extremely slow or even absent spatial convergence while considering metallic compounds was solved by suggesting the original technique, based on reciprocal space framework.
Using fundamental Fourier transform-inspired interconnection between suggested technique and traditional spatial representation, we shed light on symmetry breaking in \textit{bcc} Fe on the level of orbitally decomposed total exchange surrounding.

%% file: Introduction.tex
\section{Introduction}
\label{intro}


The interest of the world's physical community in the field of magnetic phenomena is gradually, but surely shifting towards the microscopic language of description.
It appears well-reasoned in the context of modern technological trends towards miniaturization and energy-saving, where searching of real substitution for traditional electronic semiconductor components becomes a task to be treated now.
Indeed, well-known Moore's law~\cite{moores_law_1965, moores_law_1975}, being empirically associated with retention the growth rates of computing devices' productivity, completely loses its relevance. Thus, further progress in computing power demands a fundamentally new quantum vision of basic elements, their interconnection, and functionality. 
The model systems and actual prototypes~\cite{DiVincenzo, nature_quantcomp_processor, Stanford_encyclopedia_quantcomp} are intensively investigated to fulfill this upgrade.
Thus, the real quantum processors available today~\cite{IBM_Quantum_Experience} are globally affirmed to be a key milestone on a road to completely quantum devices.

It seems completely natural that the problem of theoretical and experimental development of such devices implies a principal redefinition of structural components, being connected to each other by means of unified Hilbert space.
Free choice of its basis allows us to make an explanation from either localized atomic or delocalized "band-wise" point of view or their effective interposition.

The atomic option immediately actualizes the formalism of spin models, designed to capture magnetic properties on the level of distinguished regular particles. 
The conventional object to be studied is known to be an insulator - its magnetic atomic interplay could be well described by relatively short-ranged couplings~\cite{Slater_PhysRev_94_1498,Pavarini2012correl12,2012correl12_book,2011correl11_book,Goringe_1997,yu2005fundamentals}, which one could exhaustively take into account in the frame of a technically implementable numerical scheme.
It results in a wide range of magnetic phenomena~\cite{Bogdanov_skyrmions,ANDERSON_spin_liquids,high_temp_supercond,Mott_ins_Stepanov}, described on the basis of individual atoms and molecular orbitals.

Along with the methodological task to make the spin model physically relevant and numerically solvable, its adjustment to the real material is well-known to be a strictly non-straightforward procedure. Providing that investigation was started from the first-principles calculation of electronic structure~\cite{Kohn_Sham_1965}, the typical scheme is to construct a minimal model of magnetoactive electron shell by projecting onto Wannier functions~\cite{G.Wannier-original}.

Thus formulated tight-binding Hamiltonian allows one to represent the problem in the framework of Green's functions.
Alexander Liechtenstein, Mikhail Katsnelson, and co-authors in their pioneer works~\cite{Liechtenstein_1984,LKAG_1987,ANTROPOV1997336,LKAG_Katsnelson_2000,LKAG_Katsnelson_2004} demonstrated, that application of local force theorem~\cite{LocalForceTheorem_1, LocalForceTheorem_2} to the case of infinitesimal spin rotations remarkably allows one to estimate the magnetic environment of a single atom ${\cal{J}}_i$, as well as particular pairwise exchange interaction $J_{ij}$, directly on the base of on-site and inter-site Green's functions.

In this disposition, the relationship between these two approaches seems trivial:
\begin{equation}
    {\cal{J}}_i = \sum_{j \neq i} J_{ij}
    \, .
\label{calJi}
\end{equation}
However, numerical calculations of actual insulators, conducting systems, and model crystals~\cite{Igoshev_2015,Kvashnin_2016,Belozerov_2017,Kashin_2018} reveal significant discrepancies.

In this work, we conduct a comprehensive analysis of this problem.
Using the apparatus of on-site and inter-site Green's functions to estimate exchange interactions in simple model crystals, we were able to show analytically that the above discrepancy is practically inevitable when considering spin-polarized electron hopping integrals in the tight-binding Hamiltonian.
Because of this, the application of models with strictly on-site sources of magnetism (the Hartree-Fock method~\cite{hartree_1928,Fock1,Fock2,2016correl16_book} and the dynamic mean-field theory~\cite{vollhardt,Georges_Kotliar_1992,Georges_Kotliar_1996,kotliar_LDA_DMFT}) for the study of conducting materials seems to be fundamentally more preferable than LSDA-based approaches.

Important to note that such deviations are often associated with the presence of conduction electrons, which make the exchange interaction between atoms significant even at hundreds of Angstroms.
Indeed, the characteristic sinusoidal behavior of $J_{ij}$ with increasing distance indicates the presence of the RKKY mechanism~\cite{RKKY_Ruderman_Kittel,RKKY_Kasuya,RKKY_Yosida,Kvashnin_2016}.
From a theoretical point of view, this first of all means either very slow or totally absent convergence of spatial sums of pairwise exchange interactions $\sum_{j \neq i}~J_{ij}$.
In order to circumvent this problem, researchers often have to resort to artificial damping numerical tricks that ensure the convergence of such sums~\cite{Pajda_damping}.
In this case, of course, additional methodological difficulties arise in the physical validation of obtained estimates.

In this regard, we propose an original analytical technique, that allows one to reproduce a Fourier image of pairwise exchange interactions $J(\bm{q})$ from corresponding Green's functions.
Along with physical equality to $J_{ij}$ landscape, we demonstrate on the example of model crystals and \textit{bcc} iron, that this technique appears as the most reliable indicator of 
$\sum_{j \neq i}~J_{ij}$ actual convergence dynamics. 

We also emphasize that $\JqZero$ point turned out to be the only source of such convergence expectations if one considers the orbitally decomposed exchange interactions.
This feature enables us to study a problem concerning the net non-suppression of cross-atomic $\ttwogeg$ interplay, which is anticipated from the cubic point group symmetry in $d$-magnetics.
Employing Parseval's equality grants a remarkable possibility to practically examine its numerically inevitable residuality.


%% file: Method.tex
\section{Method}

The modern common practice for the reconstruction of equilibrium electronic, magnetic and other characteristics of solids almost invariably includes as a primary stage the first-principle modeling of the electronic structure~\cite{Hohenberg_Kohn_1964,Kohn_Sham_1965,GGA,Anisimov91}.
Further, the obtained numerical results are used to construct the so-called minimal model - the part of the system energy spectrum, which is decisive for the appearance of the properties in focus.
For these purposes, the most popular approach is to utilize a basis of maximally localized Wannier functions~\cite{G.Wannier-original,mlwf-review,Marzari_MLWF1997, Marzari_MLWFDisentangle2012}.

As a result, the minimal model is formalized by means of its Hamiltonian.
The most convenient frame is known to be the tight-binding approximation~\cite{Goringe_TB}, where it can be written as a matrix function, that characterizes both pairwise atomic couplings and on-site electron energies. Thus if we express how atom $i$ of the unit cell with translation $\bm{T} = 0$ interacts with the atom $j$ of the unit cell with translation $\bm{T}$, we write:
\begin{equation}
    \big[
    H^{\sigma \sigma'}(\bm{T})
    \big]_{ij} 
    = 
    t_{ij}^{\sigma \sigma'}
    + 
    \varepsilon_{i}^{\sigma} \, 
    \delta_{\sigma \sigma'}\, \delta_{ij} \, 
    \delta_{\bm{T} 0} \, ,
\label{hamilt}
\end{equation}
where $t_{ij}^{\sigma \sigma'}$ is the electron hopping matrix between the atoms' different orbitals, $\varepsilon_{i}^{\sigma}$ - energy matrix of orbital electrons, $\sigma$ and $\sigma'$ - spin indices, $\delta$ - Kronecker delta. Square brackets reflect the important fact that in practical calculations the Hamiltonian matrix has the size of the unit cell - hence the interatomic level is included as the corresponding matrix sectors $ij$.

In this study, it is assumed that the source of magnetism fully contains in partially filled electron orbitals. Therefore, they are to be captured in the minimal model.
Since the focus of our attention is on conducting magnets of the Heisenberg type, we do not consider site-to-site electron hopping with spin-flip:
\begin{equation}
      H^{\sigma \sigma'}(\bm{T}) = H^{\sigma}(\bm{T}) \, \delta_{\sigma \sigma'} \, .
\label{Hamilt_only_one_Spin_index}
\end{equation}
This essentially splits the model into two spin subsystems to be operated individually.
Thus, Hamiltonian (\ref{Hamilt_only_one_Spin_index}) in the reciprocal space reads:
\begin{equation}
      H^{\sigma}(\bm{k}) = \sum_{\bm{T}} \, H^{\sigma}(\bm{T}) \cdot \regexp(i \bm{k} \bm{T})
      \, .
\end{equation}
It allows one to construct $\bm{k}$-dependent Green's function:
\begin{equation}
      \GkE = \big\{ \energy - H^{\sigma}(\bm{k}) \big\} ^{-1} \, ,
\label{GreenEK}
\end{equation}
where $\energy$ is the spectrum sweep energy in the diagonal matrix form. Then we assemble its on-site and inter-site versions:
\begin{equation}
    G^{\sigma}_{i} = 
    \frac{1}{N_{\bm{k}}} 
    \sum_{\bm{k}} 
    \big[ \GkE \big]_{ii}  \, ,
\label{Gi-sigma}
\end{equation}
\begin{equation}
    G^{\sigma}_{ij} = 
    \frac{1}{N_{\bm{k}}} 
    \sum_{\bm{k}} 
    \big[ \GkE \big]_{ij}
    \cdot 
    \regexp(-i \bm{k} \bm{T}_{ij})  \, ,
\label{Gij-sigma}
\end{equation}
where $\bm{T}_{ij}$ is the translation vector connecting the cells of the $i$ and $j$ atoms, $N_{\bm{k}}$ is the number of Monkhorst-Pack grid points~\cite{Monkhorst_Pack}, $\energy$ as an argument is omitted for brevity.


\subsection{Isotropic exchange interactions}

Further, to estimate the picture of isotropic exchange interactions, 
we should perform a mapping of the original electronic model onto the effective spin model:
\begin{equation}
      {\cal{H}} = 
      -
      \sum_{ij} 
      J_{ij} \,\,
      \bm{e}_i \cdot \bm{e}_j
      \, ,
\label{SpinHamNonPerturbed}
\end{equation}
where $\bm{e}_i$ is the unit vector of "classically" approximated spin and each couple is taken twice.

For this purpose the initial configuration is assumed to be purely ferromagnetic:
$\bm{e}_i = (0, 0, 1)$.
Let us then consider an infinitesimal spin rotations on an angle $\bm{\delta \phi} = \delta \phi \cdot \bm{n}$, where $\bm{n} = (n_x, n_y, 0)$ is the axis direction.
Thus caused energy perturbation one can describe by calculating the second variation of Eq.~(\ref{SpinHamNonPerturbed}):
\begin{equation}
      \delta^2 {\cal{H}} = 
      -
      \sum_{ij} 
      J_{ij} \,
      \Big[
            \bm{\delta^2 e_i} \cdot \bm{e_j}
                  +
            2 \; \bm{\delta e_i} \cdot \bm{\delta e_j}
                  +
            \bm{e_i} \cdot \bm{\delta^2 e_j}
      \Big]
      \, ,
\label{SpinEnergySecondVariation}
\end{equation}
where 
\begin{align}
\begin{split}
      \bm{\delta e_i} = 
      [ \bm{\delta \phi_i} \times \bm{e_i} ] &= 
      (\delta \phi_i^y , \, -\delta \phi_i^x , \, 0) 
      \, , \\
      \bm{\delta^2 e_i} = 
      [ \bm{\delta \phi_i} \times \bm{\delta e_i} ] &= 
      -( 0  , \,  0  , \, \delta^2 \phi_i^x + \delta^2 \phi_i^y )
      \, .
\end{split}
\end{align}
Thereby, Eq.~(\ref{SpinEnergySecondVariation}) takes the final form:
\begin{align}
\begin{split}
      \delta^2 {\cal{H}} = 
            \sum_{i}
                  \Big\{
                        \sum_{j} J_{ij}
                        &\cdot
                        \big(
                              \delta^2 \phi_i^x + \delta^2 \phi_i^y 
                        \big)
                  \Big\}
                        + \phantom{AA} \\
          + \sum_{j}
                  \Big\{
                        \sum_{i} J_{ij}
                        &\cdot
                        \big(
                              \delta^2 \phi_j^x + \delta^2 \phi_j^y 
                        \big) 
                  \Big\}
                        + \phantom{AA} \\
          + \sum_{ij}
                  \Big\{
                        - 2 \, J_{ij}
                        &\cdot
                        \big(
                              \delta \phi_i^x \, \delta \phi_j^x
                                          + 
                              \delta \phi_i^y \, \delta \phi_j^y
                        \big)
                  \Big\}
            \, .
\label{FinalSpinVariation}
\end{split}
\end{align}
      

As the next step we should employ the local force theorem~\cite{LocalForceTheorem_1, LocalForceTheorem_2, Lichtenstein2013correl13} to the electronic Hamiltonian, Eq.~(\ref{Hamilt_only_one_Spin_index}). 
According to this theorem, the total energy variation $\delta \ElecHamE$, caused by small perturbation from the ground state of the system, could be represented as the sum of one-particle energy changes of the occupied states, with ground state potential kept fixed. In terms of first order perturbations we write for charge and spin densities~\cite{LKAG_1987}:
\begin{equation}
    \delta \ElecHamE =
    \int_{-\infty}^{\energy_{F}} 
    \energy \cdot \delta \tilde{n}(\energy)
    \, \dE 
                     =
    \energy_{F} \cdot \delta Z 
    -
    \int_{-\infty}^{\energy_{F}} 
    \delta \tilde{N}(\energy)
    \, \dE 
                     =
    -
    \int_{-\infty}^{\energy_{F}} 
    \delta \tilde{N}(\energy)
    \, \dE  
    \, ,
\end{equation}
where 
$\tilde{n}(\energy) = \mathrm{d} \tilde{N}(\energy) / \mathrm{d} \energy$ is the density of electron states, 
$\tilde{N}(\energy)$ its integrated version,
$\energy_{F}$ is the Fermi energy,
$\delta Z$ is the change of total number of electrons, being zero if we consider magnetic excitation case.

Assuming $H$ and $G$ to be short spinor notation of the Hamiltonian, Eq.~(\ref{Hamilt_only_one_Spin_index}) and Green's function, Eq.~(\ref{Gi-sigma}), for $\tilde{n}(\energy)$ one can write
\begin{equation}
      \tilde{n}(\energy) =
      - \frac{1}{\pi} \,
      \mathrm{Im} \,
      \mathrm{Tr_{L , \sigma}}
      [ G ] 
      \, ,
\end{equation}
which leads to the following expression for $\delta \tilde{N}(\energy)$:
\begin{equation}
      \delta \tilde{N}(\energy) =
      \frac{1}{\pi} \,
      \mathrm{Im} \,
      \mathrm{Tr_{L , \sigma}}
      [ \delta H \, G ]
      \, ,
\end{equation}
where 
$\mathrm{Tr_{L \, \sigma}}$ denotes the trace over orbital ($\mathrm{L}$) and spin ($\mathrm{\sigma}$) indices.

Consequently, the second variation of total energy could be expressed as
\begin{equation}
      \delta^2 \ElecHamE =
      - \frac{1}{\pi}
      \int_{-\infty}^{\energy_{F}}
      \mathrm{Im} \,
      \mathrm{Tr_{L , \sigma}} 
          \big[ 
            \delta^2 H \, G
            + 
            \delta H \, G \, \delta H \, G 
          \big]
      \, \dE  
      \, .
\label{TheSecondVariationE}
\end{equation}

In order to consider the spin rotation by $\bm{\delta \phi}$ on the level of electron model, we should introduce the corresponding operator:
\begin{equation}
      \hat{U} = \regexp 
                \Big(
                       i \,\, \frc{1}{2} \,\,
                              \bm{\delta \phi} \cdot \hat{\bm{\sigma}}
                \Big)
      \, ,
\end{equation}
where 
$\hat{\bm{\sigma}} = (\hat{\sigma}_x, \, \hat{\sigma}_y, \, \hat{\sigma}_z)$ are Pauli matrices.
Providing that $\bm{\delta \phi}$ is small, one can perform the expansion:
\begin{equation}
      \hat{U}
            \approx
                  1 
                  + 
                  i \,\, \frc{1}{2} \,\,
                              \bm{\delta \phi} \cdot \hat{\bm{\sigma}}
                  -
                  \frc{1}{8} \,\,
                             ( \bm{\delta \phi} \cdot \hat{\bm{\sigma}} )^2
                  \, .
\end{equation}
Then, being applied to the Hamiltonian, Eq.~(\ref{Hamilt_only_one_Spin_index}), this operator generates the first and the second variation for atom $i$ as follows:
\begin{equation}
      \delta H_{ii} =
      \frac{\Delta_i}{2}
      \,
      \bigg\{
            i \,
            \delta \phi^x_i
            \cdot
            \begin{pmatrix}0 & 1\\ -1 & 0 \end{pmatrix}
                  +
            \delta \phi^y_i
            \cdot
            \begin{pmatrix}0 & 1\\  1 & 0 \end{pmatrix}
      \bigg\}
      \, ,
\end{equation}
\begin{equation}
      \delta^2 H_{ii} =
            \frac{\Delta_i}{2}
            \begin{pmatrix} -1 & 0\\ 0 & 1 \end{pmatrix}
            \cdot
            \big(
                 \delta^2 \phi_i^x + \delta^2 \phi_i^y 
            \big)
      \, ,
\end{equation}
where intraatomic spin-splitting $\Delta_i$ is defined by:
\begin{equation}
    \Delta_{i} = 
    \big[ H^{\uparrow}(\bm{T} \shorteq 0) \big] _{ii} 
    - 
    \big[ H^{\downarrow}(\bm{T} \shorteq 0) \big] _{ii} 
    = 
    \frac{1}{N_{\bm{k}}} \sum_{\bm{k}} 
    \big[ H^{\uparrow}(\bm{k}) \big] _{ii} 
    - 
    \big[ H^{\downarrow}(\bm{k}) \big] _{ii} \, .
\label{DeltaEq}
\end{equation}

Thereby, the second variation of electron system reads:
\begin{align}
\begin{split}
      \delta^2 \ElecHamE =
      - \frac{1}{\pi}
     &\int_{-\infty}^{\energy_{F}}
      \mathrm{Im} \,
      \mathrm{Tr_{L}} 
      \,\, \times
      \\
      \times \,
      \bigg\{
			\phantom{+}
            \frc{1}{2}
           &\Big[ 
                  \sum_{i}
                  \mathrm{Tr_{\sigma}} ( \delta^2 H_{ii} \, G_i )
            \Big]
                   +  
			\frc{1}{2}
            \Big[ 
                  \sum_{j}
                  \mathrm{Tr_{\sigma}} ( \delta^2 H_{jj} \, G_j )
            \Big]
                  \,\, + \\
         + \, \frc{1}{2}
           &\Big[ 
                  \sum_{i}
                  \mathrm{Tr_{\sigma}} ( \delta H_{ii} \, G_{i} \, \delta H_{ii} \, G_{i} )
            \Big]
                   + 
            \frc{1}{2}
            \Big[ 
                  \sum_{j}
                  \mathrm{Tr_{\sigma}} ( \delta H_{jj} \, G_{j} \, \delta H_{jj} \, G_{j} )
            \Big]
                  \,\, + \\
                   +
           &\Big[ 
                  \sum_{ij , \, i \ne j}
                  \mathrm{Tr_{\sigma}} ( \delta H_{ii} \, G_{ij} \, \delta H_{jj} \, G_{ji} )
            \Big]
      \bigg\}
      \, \dE  
      \, ,
\label{SecondVariationElectronRotation}
\end{split}
\end{align}
where
\begin{align}
\begin{split}
      \mathrm{Tr_{\sigma}} ( \delta^2 H_{ii} \, G_i ) 
     &=
     -\frc{1}{2} \,\,
      \Delta_i
    ( G^{\uparrow}_i - G^{\downarrow}_i )
      \cdot
      \big(
           \delta^2 \phi_i^x + \delta^2 \phi_i^y 
      \big)
      \, ,
      \\
      \mathrm{Tr_{\sigma}} ( \delta H_{ii} \, G_{ij} \, \delta H_{jj} \, G_{ji} ) 
     &=
      \frc{1}{4} \,\,
      \Big\{
            \sum_{\sigma}
            \Delta_{i} \,
            G^{\sigma}_{ij} \,
            \Delta_{j} \,
            G^{- \sigma}_{ji}
      \Big\}
      \cdot
      \big(
            \delta \phi_i^x \, \delta \phi_j^x
                              + 
            \delta \phi_i^y \, \delta \phi_j^y
      \big)
      \, ,
\end{split}
\end{align}
and $- \sigma$ implies opposite spin direction to $\sigma$.

Finally, by matching Eq.~
(\ref{SecondVariationElectronRotation}) with Eq.~ 
(\ref{FinalSpinVariation}), we come to the following regular expressions for particular atom $i$ and couple $ij$:
\begin{align}
\begin{split}
      \delta^2 \phi_i^x + \delta^2 \phi_i^y 
      \, :  \,\,\,
            \sum_{j \ne i} J_{ij}
                  &=  
            \frac{1}{4 \pi}
            \int_{-\infty}^{\energy_{F}}
            \mathrm{Im} \,
            \mathrm{Tr_{L}} 
            \big[
                  \Delta_i
                  ( G^{\uparrow}_i - G^{\downarrow}_i )
            \big]
            \, \dE  
                        \phantom{A} -
                        \\
                        &-
            \frac{1}{8 \pi}
            \int_{-\infty}^{\energy_{F}}
            \mathrm{Im} \,
            \mathrm{Tr_{L}} 
            \Big[
                  \sum_{\sigma}
                        \Delta_{i} \,
                        G^{\sigma}_{ii} \,
                        \Delta_{i} \,
                        G^{- \sigma}_{ii}
            \Big]
            \, \dE 
            \, ,
\label{OneCenteredPart}
\end{split}
\end{align}
\begin{equation}
      \delta \phi_i^x \, \delta \phi_j^x
                        + 
      \delta \phi_i^y \, \delta \phi_j^y
      \, :  \,\,\,
      J_{ij} =
            \frac{1}{8 \pi}
            \int_{-\infty}^{\energy_{F}}
            \mathrm{Im} \,
            \mathrm{Tr_{L}} 
            \Big[
                  \sum_{\sigma}
                        \Delta_{i} \,
                        G^{\sigma}_{ij} \,
                        \Delta_{j} \,
                        G^{- \sigma}_{ij}
            \Big]
            \, \dE 
            \, .
\label{TwoCenteredPart}
\end{equation}


It is clearly seen, that ${\cal{J}}_i$, Eq.~(\ref{calJi}), could be found both by Eq.~(\ref{OneCenteredPart}) and Eq.~(\ref{TwoCenteredPart}), which thereafter present singe and pairwise infinitesimal spin rotations, correspondingly, as the source of energy perturbations.

However, numerical calculations reveal principal discrepancy between these two approaches. 
The situation appears even more vague if one deals with metallic systems, since in that case long-range exchange interactions strongly hamper the convergence of the real space $J_{ij}$ composition. 
Thus, in this work we propose the original technique for numerical reconstruction of the extremely delocalized Heisenberg magnetic picture, based on reciprocal space framework.


\subsection{ J(\texorpdfstring{$\bm{q}$}{TEXT}) }

The mapping of all found $J_{ij}$ values to the reciprocal space is a well-known tool that allows one to study the spin waves' dispersion spectra and to estimate the spin stiffness constants as well as the Dzyaloshinsky-Moriya interaction~\cite{Kashin_2018,Kashin_CrI3,Mazurenko_DMI_2021,MK-DMI-2009,our-nphys-sign-DMI}.
Its traditional application implies a redefinition of the basic structural units of the considered model.
Now, atoms with the same serial number in the cell form magnetic sublattices (we will denote them as the corresponding number with a tilde).
Hence the elements of $[J(\bm{q})]_{\sublati \sublatj}$ matrix acquire the meaning of the sublattices' interaction intensity:

\begin{equation}
    \big[
    J(\bm{q})
    \big]_{\sublati \sublatj} = 
    \sum_{\bm{T}_{ij}} J_{ij} \cdot \regexp(i \bm{q} \bm{T}_{ij}) \, .
\label{Jq_Fourier}
\end{equation}

However, due to impossibility of covering all the significant $J_{ij}$'s, a new expression is required to reconstruct $J(\bm{q})$ directly from the electron Green's functions.

By substituting Eq.~(\ref{TwoCenteredPart}) into Eq.~(\ref{Jq_Fourier}), we derive:
\begin{equation}
    \big[
    J(\bm{q})
    \big]_{\sublati \sublatj} = 
    \frac{1}{8 \pi} 
    \int_{-\infty}^{\energy_{F}} 
    \mathrm{Im} \,
    \mathrm{Tr_{L}} 
    \bigg( 
    \sum_{\sigma} 
    \sum_{\bm{k} \bm{k}'} 
    {\cal{A}}^{\sigma}_{ij} (\bm{k}) \cdot 
    {\cal{A}}^{-\sigma}_{ji} (\bm{k}') \cdot 
    {\cal{B}}_{ij} ( \bm{k}' - \bm{k} + \bm{q} )
    \bigg) 
    \, \dE \, ,
\label{Ilya_Jq_via_A}
\end{equation}
where
\begin{equation}
    {\cal{A}}^{\sigma}_{ij} (\bm{k}) = 
    \frac{1}{N_{\bm{k}}}
    \,
    \Delta_{i} 
    \big[ \GkE \big]_{ij} \, ,
\end{equation}
\begin{equation}
    {\cal{B}}_{ij} ( \bm{k}' - \bm{k} + \bm{q} ) = 
    \sum_{{\bm{T}}_{ij}} 
    \regexp 
    \big(
    i \, \{ \bm{k}' - \bm{k} + \bm{q} \} {\bm{T}}_{ij}
    \big)  \, .
\end{equation}

Taking into account that 
${\cal{B}}_{ij} (\bm{k}' - \bm{k} + \bm{q}) = 
N_{\bm{k}} \cdot \delta(\bm{k}' - \bm{k} + \bm{q})$, 
we can state the final expression:
\begin{equation}
    \big[
    J(\bm{q})
    \big]_{\sublati \sublatj} = 
    \frac{N_{\bm{k}}}{8 \pi} 
    \int_{-\infty}^{\energy_{F}} 
    \mathrm{Im} \,
    \mathrm{Tr_{L}} 
    \bigg( 
    \sum_{\sigma} 
    \sum_{\bm{k}} 
    {\cal{A}}^{\sigma}_{ij}  (\bm{k}+\bm{q}) 
    \cdot 
    {\cal{A}}^{-\sigma}_{ji} (\bm{k}) 
    \bigg)
    \, \dE \, .
\label{Jq_Final_Expression}
\end{equation}

The basic feature of this result is that derivation was performed in the framework of pairwise infinitesimal spin rotations technique. In this context we highlight the remarkable usefulness of 
$\big[ \JqZero \big]_{\sublati \sublatj}$.
Indeed, Eq.~(\ref{Jq_Fourier}) in this case illustrates the straightforward possibility to reproduce ${\cal{J}}_i$ as:
\begin{equation}
      {\cal{J}}_i = 
		\sum_{\sublatj} 
		\big[ \JqZero \big]_{\sublati \sublatj}
		-
		\frac{1}{N_{\bm{q}}}
		\sum_{\bm{q}}
		\sum_{\sublatj}
        \big[ J(\bm{q}) \big]_{\sublati \sublatj}
	    \, .
\label{calJasJqZeroMinusAverageJq}
\end{equation}

Extremely important to note, that Fourier transform-driven relation between 
$J_{ij}$ 
and 
$\big[ J(\bm{q}) \big]_{\sublati \sublatj}$
demands us to formally define on-site parameter $J_{ii}$:
\begin{equation}
      J_{ii} = 
		\frac{1}{N_{\bm{q}}}
		\sum_{\bm{q}}
		\sum_{\sublatj}
        \big[ J(\bm{q}) \big]_{\sublati \sublatj}
	    \, .
\label{JiiASAverageJq}
\end{equation}
The valuable option is that $J_{ii}$ could be equally calculated by Eq.~(\ref{TwoCenteredPart}). 
It reveals the lowest cost scheme of ${\cal{J}}_i$ estimation, based only by two characteristics, being free of real spatial convergence problem and $\bm{q}$-grid density factor.

Thus, the prime novelty of this theoretical approach actualizes itself in new prospects of making a direct comparison between single and pairwise infinitesimal spin rotation techniques on the level of representative scalars. If we denote the first term of Eq.~(\ref{OneCenteredPart}) as
\begin{equation}
    {\cal{F}}_i = 
   \frac{1}{4 \pi}
   \int_{-\infty}^{\energy_{F}} 
   \mathrm{Im} \,
   \mathrm{Tr_{L}}
   \big[ 
   \Delta_{i} ( G_{i}^{\uparrow} - G_{i}^{\downarrow} )
   \, \big] 
   \, \dE \, ,
\label{calF_Definition}
\end{equation}
the perfect correspondence of the techniques is appeared to satisfy the equation:
\begin{equation}
    {\cal{F}}_i = 
    \sum_{\sublatj} 
    \big[ \JqZero \big]_{\sublati \sublatj} \, .
\label{Prime_InSpRot_Equality}
\end{equation}
For metallic systems this way manifests itself as the only one available, owing to $\sum\limits_{j \neq i} J_{ij}$ convergence problem.

We also note the self-sufficiency of reciprocal space consideration. In addition to well-known application, intended to reconstruct the spin-wave dispersion spectra~\cite{Kashin_CrI3, PhysRevB.92.144407,Okumura_2019,FUKAZAWA2019296,Solovyev_1999,Solovyev_1998,Modarresi_2019}, it is worthy to state the natural ability to find any $J_{ij}$ by inverse Fourier transform of $J(\bm{q})$, calculated on some $\bm{q}$-grid:
\begin{equation}
    J(\bm{T}_{ij}) = 
    \frac{1}{N_{\bm{q}}} 
    \sum_{\bm{q}} 
    \big[ J(\bm{q}) \big]_{\sublati \sublatj} 
    \cdot
    \regexp(-i \bm{q} \bm{T}_{ij}) \, .
    \label{JTij_from_Jq}
\end{equation}
where $J(\bm{T}_{ij})$ is the exchange interaction matrix for the corresponding pair of crystal's unit cells. Important to add that in this expression $\bm{k}$-grid and $\bm{q}$-grid have absolutely no constrains of being equal to each other. Further we will practically elaborate this question.

%% file: Exchange_Surrounding_Problem.tex
\section{Exchange Surrounding Problem}

Let us recall that in numerical calculations there is a problem of inconsistency between approaches based on a single and paired infinitesimal spin rotation. In context of our analytical derivation, the divergence could be readily enumerated as the residuality of Eq.~(\ref{Prime_InSpRot_Equality}):
\begin{equation}
    {\cal{D}}_i = 
    \sum_{\sublatj} 
    \big[ \JqZero \big]_{\sublati \sublatj}
    -
    {\cal{F}}_i \, .
\end{equation}


\subsection{Toy Model}

\begin{figure}[!h]
\centering
\includegraphics[width=0.99\columnwidth]{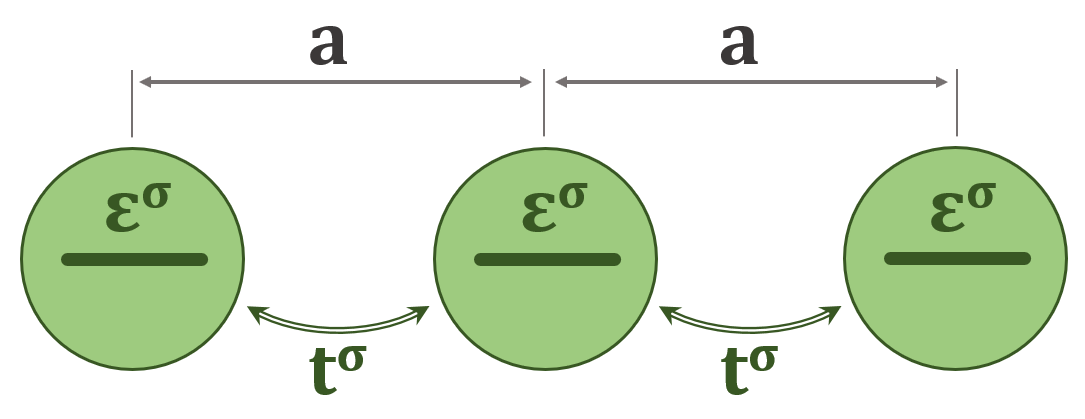}
(a)

\includegraphics[width=0.99\columnwidth]{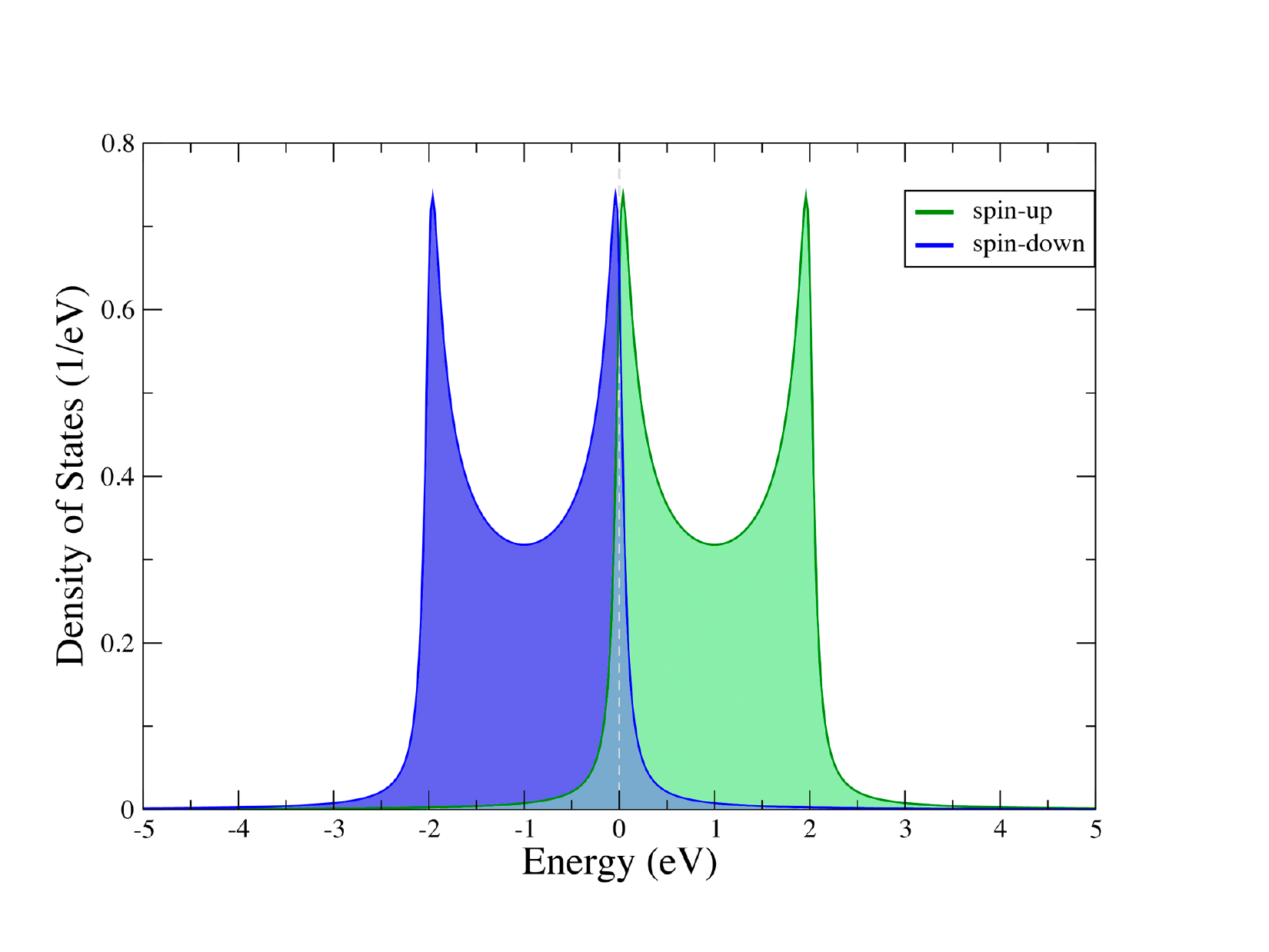}
(b)
\caption{
(a) 
Schematic representation of the toy model.
(b) 
Toy model densities of states presented for both spin channels separately. Fermi level is set at zero.}
\label{fig:toy_model}
\end{figure}

As the simplest theoretical object, we consider a \textit{toy model} of one-dimensional extended chain of identical single orbital atoms. 
This type of models could be applied to reconstruct the gapped state of real materials if the basic structure appears highly entangled~\cite{PhysRevB.81.125131}.
It is assumed to have a period of $a$ and one atom per unit cell.
The atom in the "central" cell ($T = 0$) is denoted by $i$, while all other atoms are indexed by $j$ and form the exchange surrounding of $i$ atom (Figure~\ref{fig:toy_model} (a)).
The Hamiltonian of such a crystal we express as on-site energy
$H^{\sigma}(T \, \shorteq 0) = \varepsilon^{\sigma}$ 
and the nearest neighbors hoppings 
$H^{\sigma}(T \, \shorteq \pm a) = t^{\sigma}$. 

For the numerical investigation we set the parameters as following: 
$a~=~1$~\AA, 
$\varepsilon^{\uparrow}~=~-1$~eV, 
$\varepsilon^{\downarrow}~=~1$~eV, 
$E_{F}~=~0$~eV.

\begin{figure}[!h]
\includegraphics[width=\columnwidth]{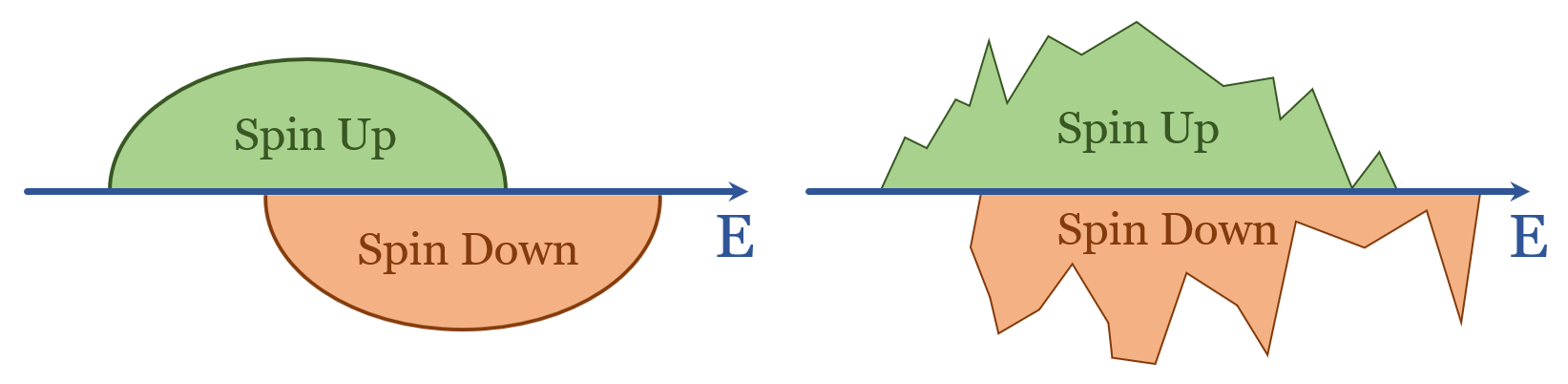}
\caption{Two cases of spin-polarized electron structure: simple band shift (\textit{left}) and qualitative band difference (\textit{right}).}
\label{fig:Models_of_Magnetism}
\end{figure}

At the first step we assume 
$t^{\uparrow} = t^{\downarrow} = -0.5$~eV.
This configuration corresponds to the simplest case of spin-polarized electron structure, where two spin subsystems differs only by a simple shift along the energy axis (Figure~\ref{fig:Models_of_Magnetism}, left).
Figure~\ref{fig:toy_model} (b) demonstrates that the model is designed to artificiality enhance the system's metallicity by boosting the DOS intensity near the Fermi level.

Important to note that the ground state of such system is known to be antiferromagnetic. And despite our approach being designed to provide the information about stability of the crystal in its tight-binding representation (collinear spin ordering), we assume our toy model relevant for fundamental analysis. It could be validated by readily available option to consider two atoms in the unit cell with opposite signs of the intraatomic spin splitting $\Delta$, which readily states AFM to be energy-lowing configuration with the picture of $J_{ij}$ kept unmodified.

The main numerical result obtained for this configuration is that 
${\cal{D}} = 0$. Moreover, it appeared valid for any particular value of 
$t^{\uparrow} = t^{\downarrow}$ 
and any settings of all other parameters. 
Therefore, we state \textbf{\textit{the absence}} of approaches divergence. Hereinafter for the crystals with one atom per unit cell for the sake of brevity we omit the single-standing index of the "central" atom and double indices of corresponding sublattice.

\begin{figure}[!h]
\includegraphics[width=0.465\columnwidth]{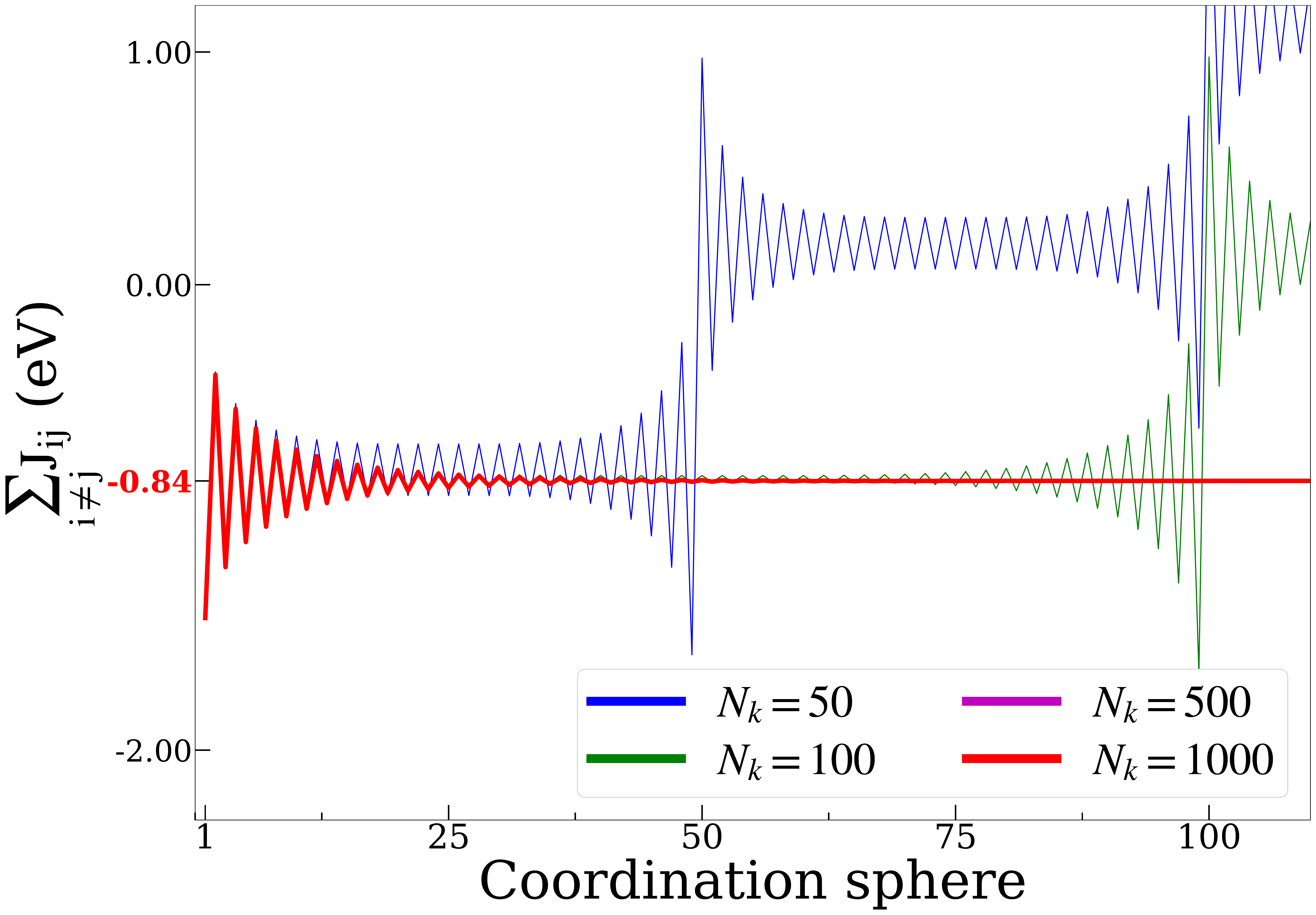}
\ 
\includegraphics[width=0.49\columnwidth]{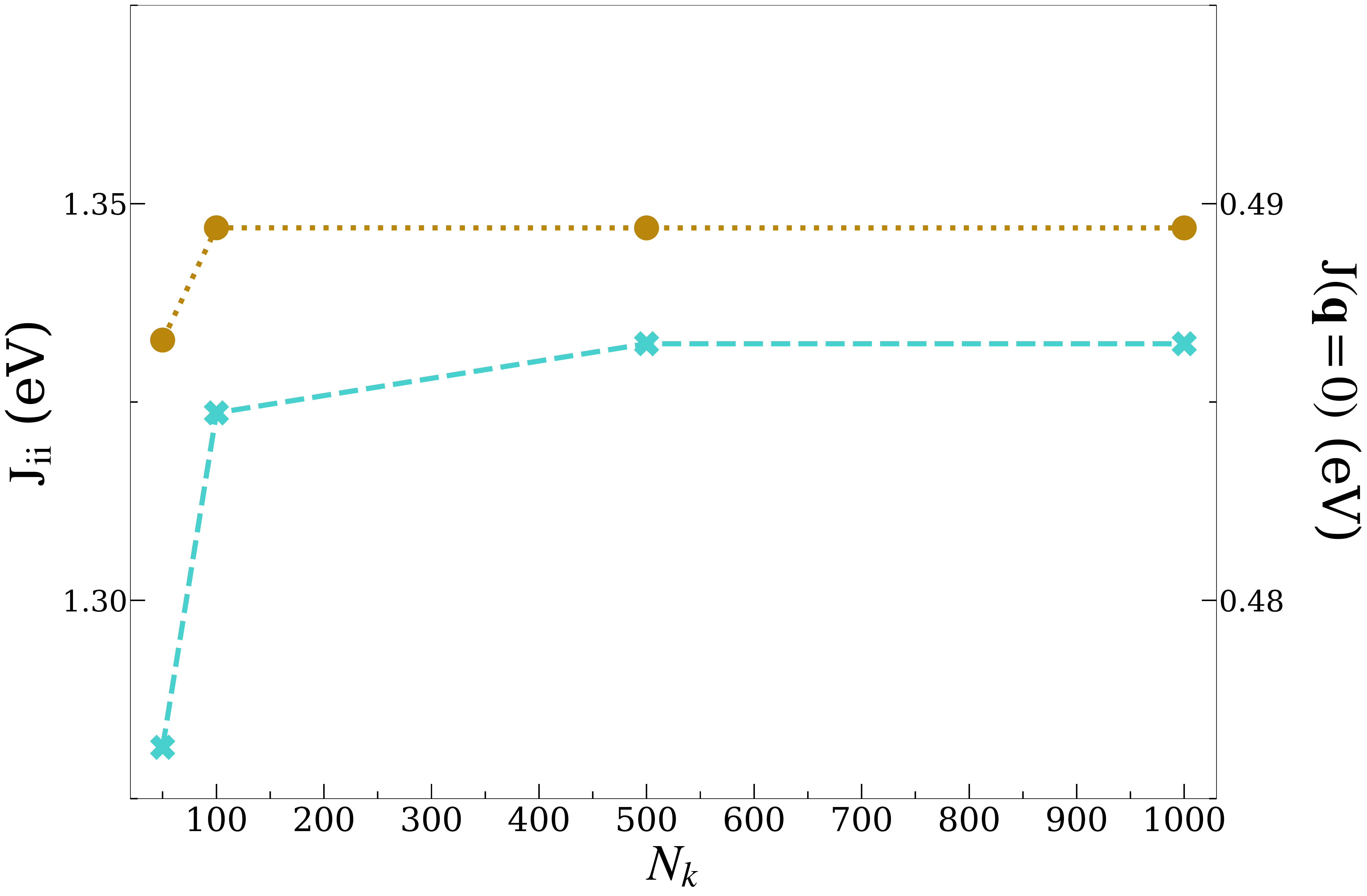}
\caption{
(\textit{Left chart}) 
Convergence dynamics of 
$\sum\limits_{j \neq i} J_{ij}$ as more distant neighbors are taken into account.
Toy model Hamiltonian's $\bm{k}$-point mesh is 
$N_{\bm{k}} \times 1 \times 1$.
(\textit{Right chart}) 
$\JqZero$ (gold) and $J_{ii}$ (turquoise) as the function of $N_{\bm{k}}$.}
\label{fig:toy_model_Jsum_convergence}
\end{figure}

In this framework we can clearly demonstrate the advantages of $J(\bm{q})$ formalism developed in our study.
Figure~\ref{fig:toy_model_Jsum_convergence} (\textit{Left chart}) shows the dynamics of the $\sum\limits_{j \neq i} J_{ij}$ spatial sum convergence as the contributions from more distant atoms are included.
It can be seen that an increase in $\bm{k}$-grid density makes the dynamics more tolerant to destabilizing high-frequency harmonics, which inevitably arise when the Hamiltonian, Eq.~(\ref{Hamilt_only_one_Spin_index}), is transformed to the reciprocal space.
In the case of one-dimensional $\bm{k}$-grid with $N_{\bm{k}} = 500$ and higher, the final value of the sum is expected to coincide with 
\begin{equation}
\JqZero - J_{ii} = {\cal{F}} - J_{ii} \, .
\end{equation}
It appears instructive to add that the value of $\JqZero$ turns out to be independent from the $\bm{q}$-grid density, in accordance with the specifics of Eq.~(\ref{Jq_Final_Expression}).

\begin{figure}[!h]
\includegraphics[width=0.475\columnwidth]{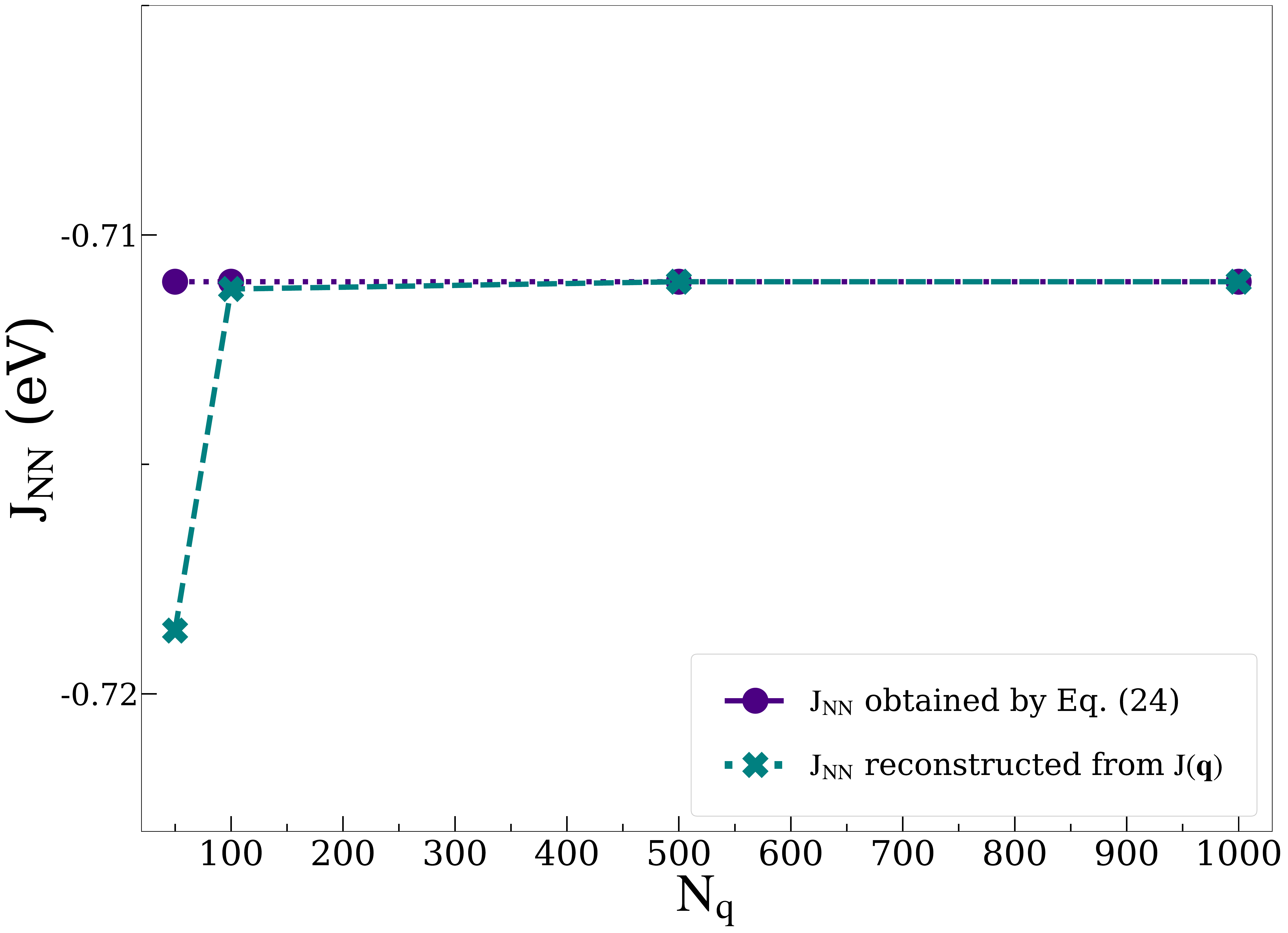}
\ 
\includegraphics[width=0.49\columnwidth]{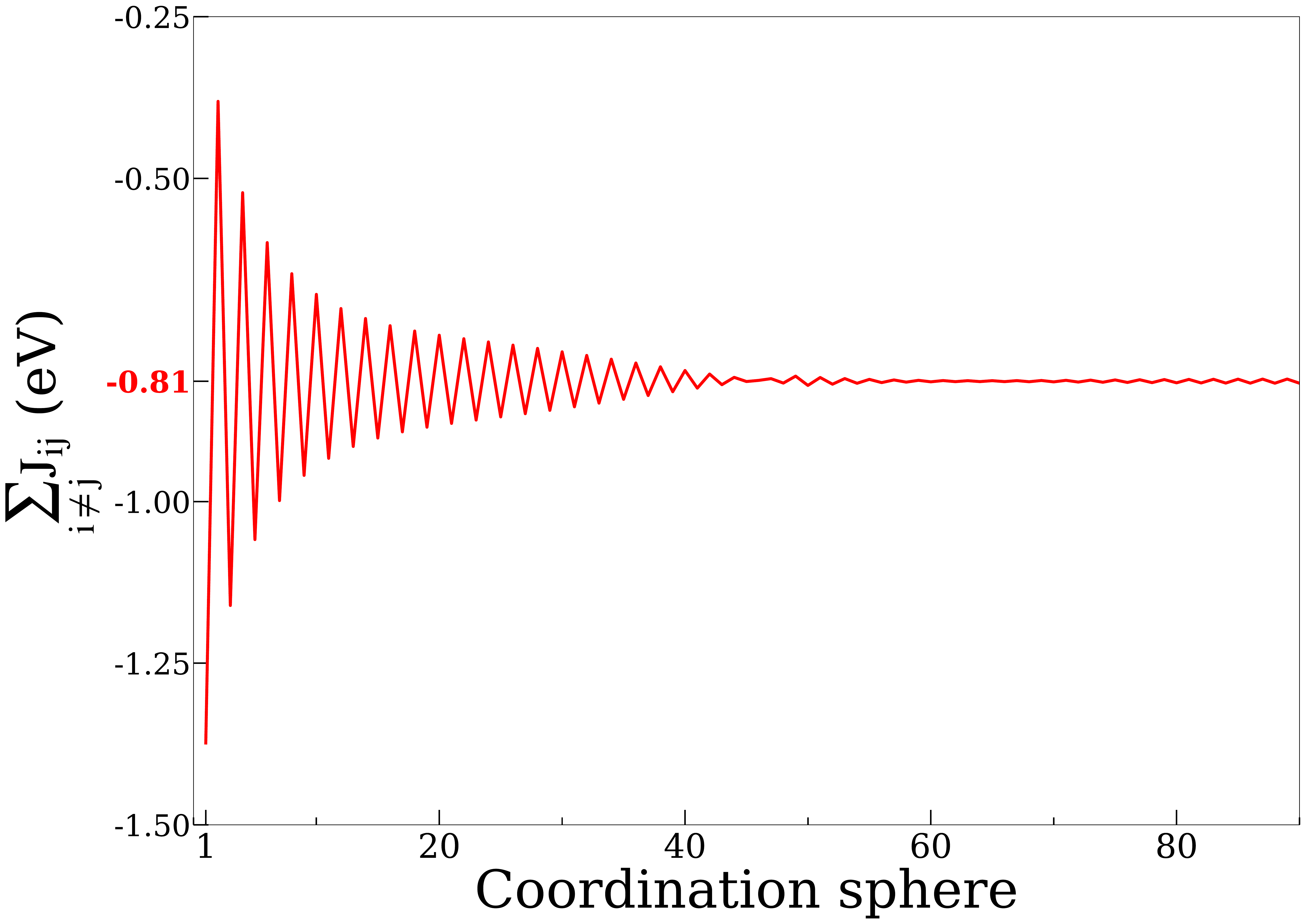}
\caption{
(\textit{Left chart}) 
Reconstructed (green) and obtained within Eq.~(\ref{TwoCenteredPart}) (purple)  values of $J_{NN}$ (between the nearest neighbors) as the function of $N_{\bm{q}}$.
(\textit{Right chart}) 
Reconstructed $J_{ij}$ on the sparse $\bm{k}$-grid ($N_{\bm{k}}$ = 50) and dense $\bm{q}$-grid ($N_{\bm{q}}~=~1000$) as the convergence dynamics of spatial sum. The expected value is estimated by Eq.~(\ref{calJasJqZeroMinusAverageJq}) as $-0.814$ eV.}
\label{fig:toy_model_J1_reconstruction}
\end{figure}

Finding all $J(\bm{q})$ along some $\bm{q}$-grid, we have an opportunity to estimate the $J_{ij}$ reconstruction accuracy, according to Eq.~(\ref{JTij_from_Jq}).
Our calculations show that if $\bm{k}$- and $\bm{q}$-grids are the same, any $J_{ij}$ can be represented by \textit{machine} precision (Figure~\ref{fig:toy_model_J1_reconstruction} (\textit{Left chart})).

It is also extremely important to consider the case of a sparse $\bm{k}$-mesh with $N_{\bm{k}}$ = 50, where the estimates of exchange interactions by Eq.~(\ref{TwoCenteredPart}) are confirmed unstable, as well as $J_{ii}$, breaking the Eq.~(\ref{JiiASAverageJq}) by means of real and reciprocal space configuration.
Computing $J(\bm{q})$ on a dense $\bm{q}$-grid with $N_{\bm{q}} = 1000$, we can observe a converging dynamics of the sum of all reconstructed $J_{ij}$ (see Figure~\ref{fig:toy_model_J1_reconstruction} (\textit{Right chart})).

In this case, the approaching value can be estimated \textit{exclusively} by Eq.~(\ref{calJasJqZeroMinusAverageJq}), but only if $J_{ii}$ is found as an extreme point of the inverse Fourier transform.
This assessment is fully performed on the base of pairwise infinitesimal spin rotations, providing guaranteed connection between individual $J_{ij}$'s and highly non-localized magnetic picture possessed by metals.

Important to add, that since increasing density of the $\bm{k}$-grid is usually associated with a significant growth of the technical requirements for computing systems, distinct approach has an another valuable advantage as it scales the  calculation time cost only, keeping Random-Access Memory demands constant.

\begin{figure}[!h]
\includegraphics[width=\columnwidth]{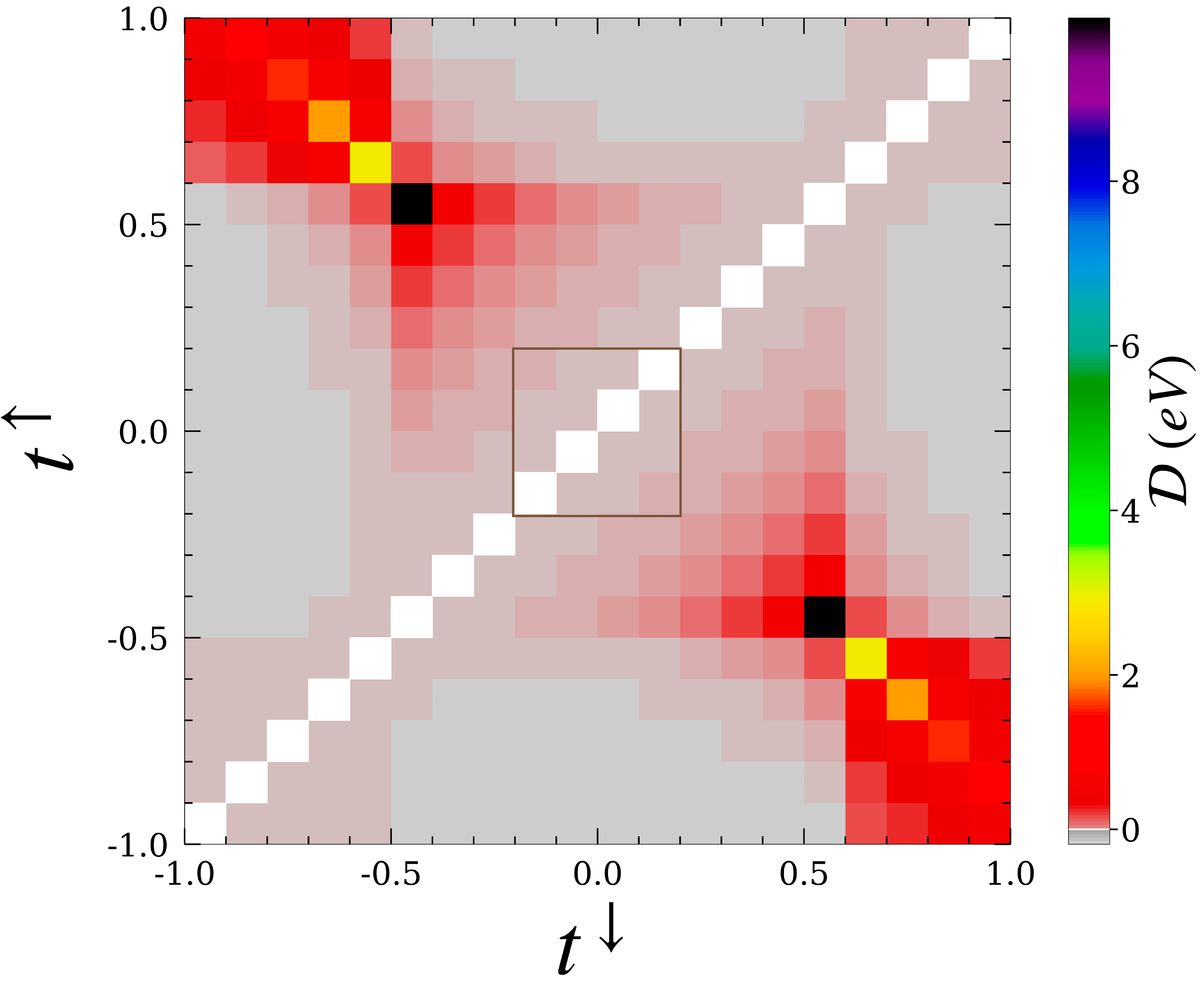}
\caption{Absolute value of ${\cal{D}}$ (in eV) as a function of $t^{\uparrow}$ and $t^{\downarrow}$ (in eV) for the considered toy model. Gold square in the middle indicates the insulator regime $t^{\sigma}~\ll~\Delta$. Note that ${\cal{D}}$ is exactly zero only if $t^{\uparrow} = t^{\downarrow}$ (white color).}
\label{fig:toy_model_Diff_2Fz_Jq0_K1000}
\end{figure}

However, the situation with the approaches consistency changes dramatically when we consider the spin-polarized case with 
$t^{\uparrow} \neq t^{\downarrow}$.
The complicated landscape of  
${\cal{D}} (t^{\uparrow}, \, t^{\downarrow})$, 
shown in Figure~\ref{fig:toy_model_Diff_2Fz_Jq0_K1000}, clearly demonstrates that the divergence turned out to be non-negligible and hard-predictable.

We highly stress that equality (\ref{Prime_InSpRot_Equality}) is restored exclusively in the $t^{\uparrow} = t^{\downarrow}$ regime.
It makes us state the fundamental principality of the magnetism model in use - strictly on-site magnetism precursor (Hartree-Fock, LDA+DMFT), which naturally inherits this regime, or spin-polarized electron gas (LSDA), which naturally does not.


Matter of fact, the general relevance of both approaches is undoubtedly kept affirmed. 
Indeed, LSDA essentially allows to capture the spin-polarized band structure from the mean-field point of view. 
This point makes it reliably applicable to isotropic and homogeneous systems, or to subsystems of more complex materials, possessed by similar properties.
Thus it accordingly gives a physically valid picture of ionization energies of atoms, binding energies of solids, bulk lattice constants, anisotropic effects etc.~\cite{PhysRevB.57.1505, Logemann_2017, PhysRevB.58.15496, PhysRevLett.76.4825, PhysRevB.53.7158}.
However, when one deals with highly inhomogeneous structures, such as strongly correlated materials~\cite{PhysRevB.92.144407}, polarized insulators~\cite{PhysRevLett.99.126405} or the systems with heavy fermions~\cite{PhysRevLett.54.1852}, low resemblance to a non-interacting electron gas produces quantitative and qualitative discrepancies between theory and experiment. For instance, we can mention systematical underestimation of the band gap~\cite{PhysRevLett.52.1830, PhysRevB.30.4734, PhysRevB.57.1505} or non-credible representation of metal-insulator transition~\cite{JETP_Anisimov, PhysRevB.78.140404, PhysRevB.78.155107}.

On the other hand, LDA+DMFT approach manifests itself as to rigorously capture electron-electron correlations at the focused area of the band structure.
Therefore, we obtain a physically motivated access to both high and low energy quasi-particle excitations and, as a consequence, partly describe high temperature properties~\cite{Panda_2014, JETP_Korotin, Gerasimov_2021} and paramagnetism~\cite{Panda_2014, Gerasimov_2021, Belozerov_2016}. Being usually applied to strongly-correlated systems~\cite{PhysRevLett.90.056401, PhysRevB.92.245135, PhysRevB.88.085112}, this method turns significantly valuable, if the magnetism precursor (or, generally, the source of the investigating property) appeared isolated in frame of well disentangled bands. Otherwise, we face both the methodical issues (double counting problem~\cite{PhysRevLett.115.196403}) and technical troubles (extra large Hamiltonian to be treated).

Summing up, we highlight the additional fundamental factor, which should be taken into account carefully if one is intended to study collective magnetic characteristics, formed by individual or pairwise atomic contributions.


\subsection{Analytical Explanation}

For a rigorous analytical substantiation of this statement, we consider a completely general case of the Hamiltonian, Eq.~(\ref{Hamilt_only_one_Spin_index}), which has \textit{spin-polarized} only on-site electron energies, whereas all hoppings are kept \textit{non-spin-polarized} (Figure~\ref{fig:Models_of_Magnetism}, left). 

For our derivation it appears principal to keep the framework of unit-cell-sized Hamiltonian (and Green's functions) matrices, implying atomic detailization incorporated. Thus we write Hamiltonian as a function of translation vector:
\begin{equation}
    H^{\sigma}(\bm{T})
    = 
    t(\bm{T}) 
    + 
    \varepsilon^{\sigma} \, 
    \delta(\bm{T}) 
    \, .
\label{Analytical_explanation_Hamiltonian}
\end{equation}
Hence in reciprocal space we got
\begin{equation}
      H^{\sigma}(\bm{k}) =
        \varepsilon^{\sigma}
        +
        {\cal{H}}(\bm{k}) \, .
\end{equation}
Then if one combines the Green's functions, Eq.~(\ref{GreenEK}):
\begin{equation}
      \big\{ \GkEdown \big\}^{-1}
       - 
      \big\{ \GkEup \big\}^{-1} 
       =
      H^{\uparrow}(\bm{k})
       -
      H^{\downarrow}(\bm{k})
       =
      \varepsilon^{\uparrow}
       -
      \varepsilon^{\downarrow}
       =
      \Delta \, ,
\end{equation}
the result essentially appears \textit{independent} from $\bm{k}$.

As the next step we express the equality (\ref{Prime_InSpRot_Equality}) at the level of on-site and inter-site Green's functions \cite{LKAG_1987}. Here on-site variant is denoted by tilde and inter-site one is given as a function of $\bm{T}$:
\begin{equation}
   {\tilde{G}}^{\uparrow} - {\tilde{G}}^{\downarrow} = 
   \sum_{\bm{T}}
   G^{\uparrow}(\bm{T})
   \cdot 
   \Delta 
   \cdot 
   G^{\downarrow}(-\bm{T}) \, .
\end{equation}
Taking into account definitions:
\begin{align}
\begin{split}
G^{\uparrow}(\bm{T}) =
   \frac{1}{N_{\bm{k}}}
   \sum_{\bm{k}} \GkEup \cdot \regexp(-i \bm{k} \bm{T})
   \\
G^{\downarrow}(-\bm{T}) =
   \frac{1}{N_{\bm{k'}}}
   \sum_{\bm{k'}} \GkprimeEdown \cdot \regexp(i \bm{k'} \bm{T})
\end{split}
   \; \; ,
\end{align}
leads us to
\begin{equation}
   {\tilde{G}}^{\uparrow} - {\tilde{G}}^{\downarrow} =
   \frac{1}{N_{\bm{k}}}
   \sum_{\bm{k}}
   \GkEup
   \cdot
   \Delta
   \cdot
   \GkEdown \, .
\label{Analytical_Exp_Aux}
\end{equation}
And finally if we write $\Delta$ for any $\bm{k}$ as:
\begin{equation}
      \Delta =
      \big\{ \GkEdown \big\}^{-1}
       - 
      \big\{ \GkEup \big\}^{-1} \, ,
\end{equation}
then Eq.~(\ref{Analytical_Exp_Aux}) turns to identity. 

Therefore, dealing with spin-polarized electron gas (Figure~\ref{fig:Models_of_Magnetism}, right), one can state ${\cal{D}}$ - according to (\ref{DeltaEq}) - proportional to statistical dispersion of
$H^{\uparrow}(\bm{k}) 
    - 
 H^{\downarrow}(\bm{k})$ 
over the 1st Brillouin zone.
In particular, for our 1D toy model we got
\begin{equation} 
        H^{\uparrow}(\bm{k}) - H^{\downarrow}(\bm{k}) =
            \Delta
            +
            2 \,
            (t^{\uparrow} - t^{\downarrow})
            \cdot
            \mathrm{cos} (k a)
\end{equation}
which clearly demonstrates that ${\cal{D}} \sim t^{\uparrow} - t^{\downarrow}$.

As a consequence, we confirm a general preference for on-site-based methods of describing magnetism if we are intended to reproduce characteristics whose definitions include the spatial sums of pairwise exchange interactions.

\input{bccFe_ExchSurr}

%% file: bccFe_ExchSurr.tex
\subsection{\textit{bcc} Fe}

To elaborate the question we consider the case of real metallic material.
As a bright representative in this paper let us analyze the canonical case of \textit{bcc} iron.
The pioneer works devoted to the theoretical study of the magnetic properties  date back to the mid 80's~\cite{LKAG_1987}.
However, scientific discussions do not subside until now~\cite{Kvashnin_2016,Cardias_2017,Wang_2010,Jacobsson_2013,Jacobsson2017ParameterisationON,magnons-jij,Turek_2006}.
In many respects, the controversy is supported precisely by the fact that the rate of convergence of spatial sums of exchange interactions is extremely low, which makes it difficult to make a reliable estimate of both the magnetic transition temperature and the spin-wave stiffness constants~\cite{Tanaka_2020,Turek_2006,magnons-jij}.

\begin{figure}[!h]
\includegraphics[width=0.49\columnwidth]{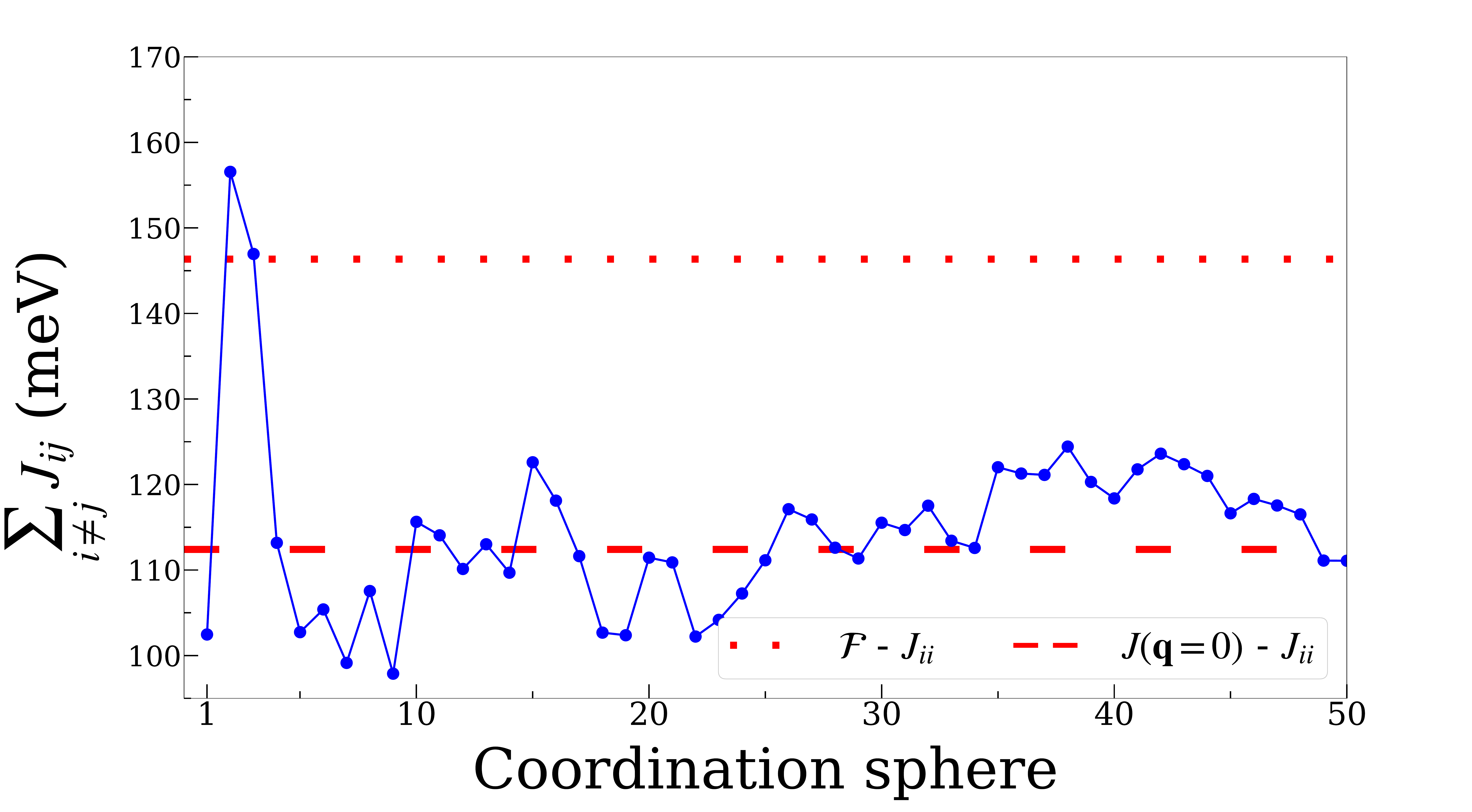}
\ 
\includegraphics[width=0.49\columnwidth]{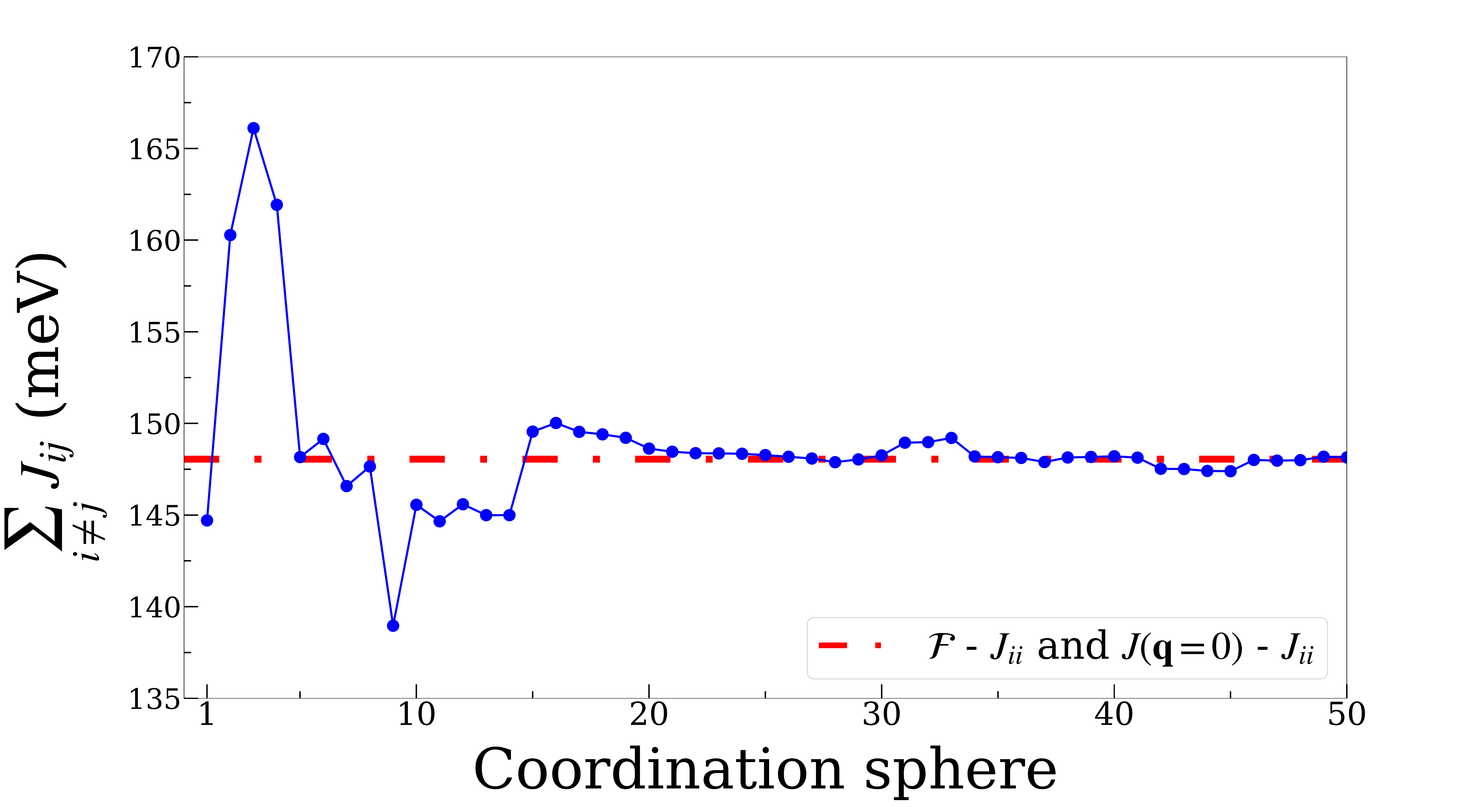}
\caption{
(\textit{Left chart}) 
LSDA-originated convergence dynamics of \bccFe $\sum\limits_{j \neq i} J_{ij}$, as more distant neighbors are taken into account.
(\textit{Right chart}) 
LDA+DMFT-originated convergence dynamics of \bccFe $\sum\limits_{j \neq i} J_{ij}$, as more distant neighbors are taken into account.}
\label{fig:bccFe_TotalJij_Convergence}
\end{figure}

To fulfill the comprehensive study of this problem we consider \bccFe magnetism from the viewpoint of spin-polarized electron gas (LSDA) and dynamical mean-field theory (LDA+DMFT).
The details of the calculations are given in the Appendix. 
Here we only mention that one atom per unit cell allows keeping the single-standing index $i$ of the "center" atom (in the cell with $\bm{T} = 0$) and double index of only sublattice omitted.
Figures \ref{fig:bccFe_TotalJij_Convergence} (\textit{Left chart}) and (\textit{Right chart}) show the dynamics of the $J_{ij}$'s convergence with the background of ${\cal{F}}~-~J_{ii}$ and $\JqZero - J_{ii}$, where $J_{ii}$ can be is equivalently found both using Eq.~(\ref{TwoCenteredPart}) and Eq.~(\ref{JiiASAverageJq}) (assuming that $\bm{k}$- and $\bm{q}$-grids are the same). All results appear in a good consistence with previous studies~\cite{LKAG_1987,LKAG_Katsnelson_2000,Kvashnin_2015_2,Belozerov_2017}.
As expected, intraatomic magnetism model, possessed by LDA+DMFT, leads the approaches based on single and pairwise spin rotations to give equal numerical results.
The divergence is observed in the LSDA case, where the value of $\JqZero - J_{ii}$ turns out to be reflecting actual dynamics of the convergence trend much more reliable.
This confirms our basic thesis that this approach inherits all the features of the pairwise spin rotation technique and does not lose methodological interconnection with it.

%% file: Orbital_Decomposition_and_Symmetry_Problem.tex
\section{Orbital Decomposition and Symmetry Problem}

In addition to the considered problem, there is also a conjuncted one, which the similar analysis tools appeared applicable to.
Its description we should initiate by the fact that the values of exchange interactions $J_{ij}$, Eq.~(\ref{TwoCenteredPart}), can be decomposed into the orbital components.
For this, it is sufficient to turn to the mathematical property of the trace operation, taken from the product of two arbitrary $N \times N$ matrices:
\begin{equation}
    \mathrm{Tr} [ X \cdot Y ] = 
    \sum_{m = 1}^{N} \sum_{l = 1}^{N} X_{ml} * Y_{lm} = 
    \sum_{m,\,l = 1}^{N} Z_{ml} .
\end{equation}
In our case, the contribution from the interaction of orbital $\alpha$ (atom $i$) with orbital $\beta$ (atom $j$) can be found as:
\begin{equation}
    \begin{split}
        \{ J_{ij} \}^{\alpha \beta} = 
        \frac{1}{8 \pi} \,
        \mathrm{Im} 
        \int_{-\infty}^{E_F} 
        \Big\{ 
        [ \Delta_{i} &\cdot G^{\uparrow}_{ij} ]^{\alpha \beta} \cdot 
        [ \Delta_{j}  \cdot G^{\downarrow}_{ji} ]^{\beta \alpha} 
        + \\
     &+ [ \Delta_{i}  \cdot G^{\downarrow}_{ij} ]^{\alpha \beta} \cdot 
        [ \Delta_{j}  \cdot G^{\uparrow}_{ji} ]^{\beta \alpha} 
        \Big\} 
        \, \dE \, . 
    \end{split}
\end{equation}
Thereafter $J_{ij} = \sum\limits_{\alpha \beta} \{ J_{ij} \}^{\alpha \beta}$.
It is indicated in works~\cite{Kvashnin_2016,Cardias_2017,Szilva_2017} that the decomposition thus constructed should have the symmetry properties of the crystal.
In particular, when considering the $d$-magnetism of real materials, the cubic point group symmetry sets the expectation of the complete suppression of the contributions from the cross interaction of the $t_{2g}$ and $e_g$ orbitals upon the spatial summation of all $J_{ij}$ around a particular atom $i$.
However, actual numerical calculations of conducting materials do not justify such expectations.

\begin{figure}[!h]
\centering
\begin{minipage}[b]{0.47\linewidth}
\centering
\includegraphics[width = 0.99 \textwidth]{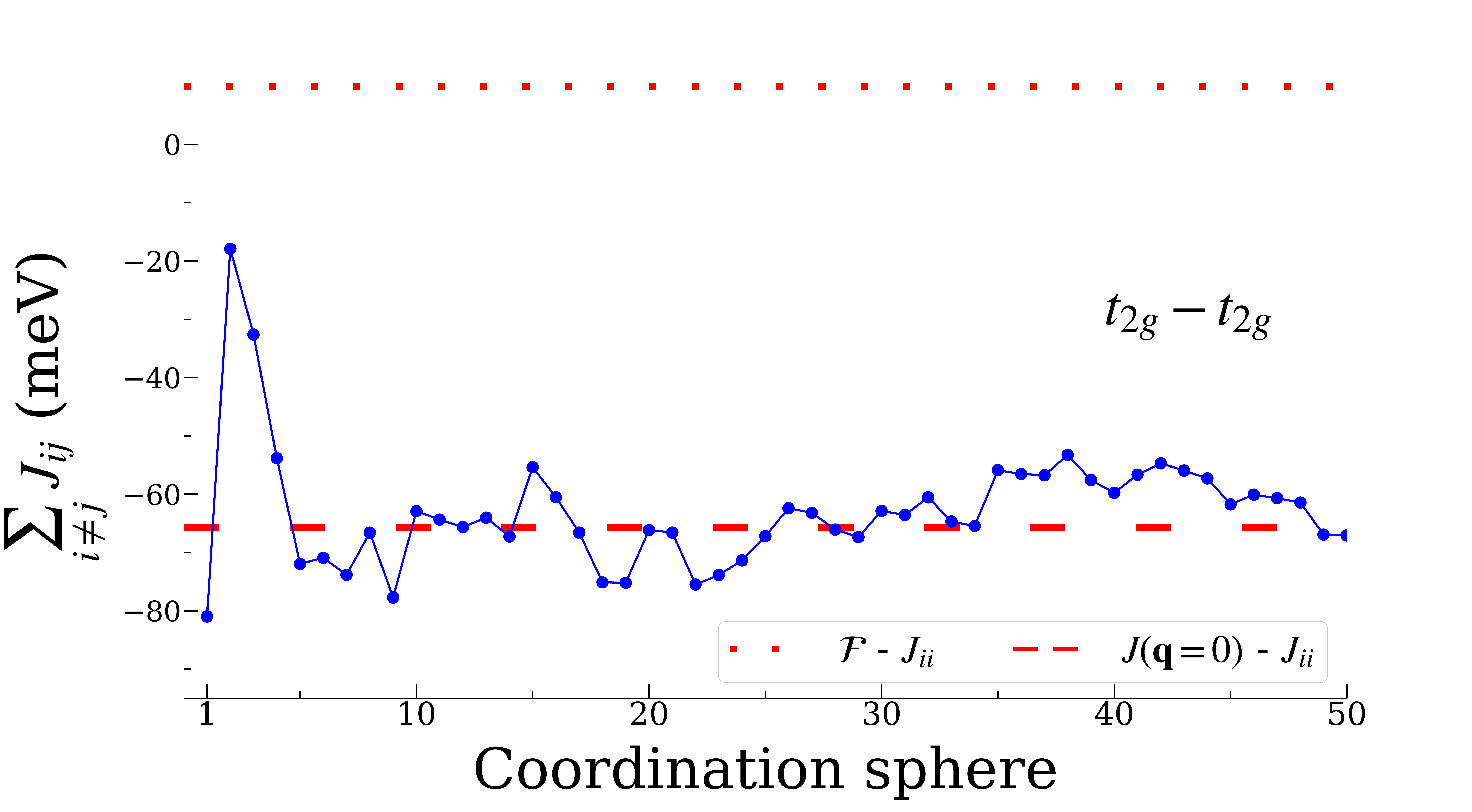}

(a)
\end{minipage}
\begin{minipage}[b]{0.49\linewidth}
\centering
\includegraphics[width = 0.99 \textwidth]{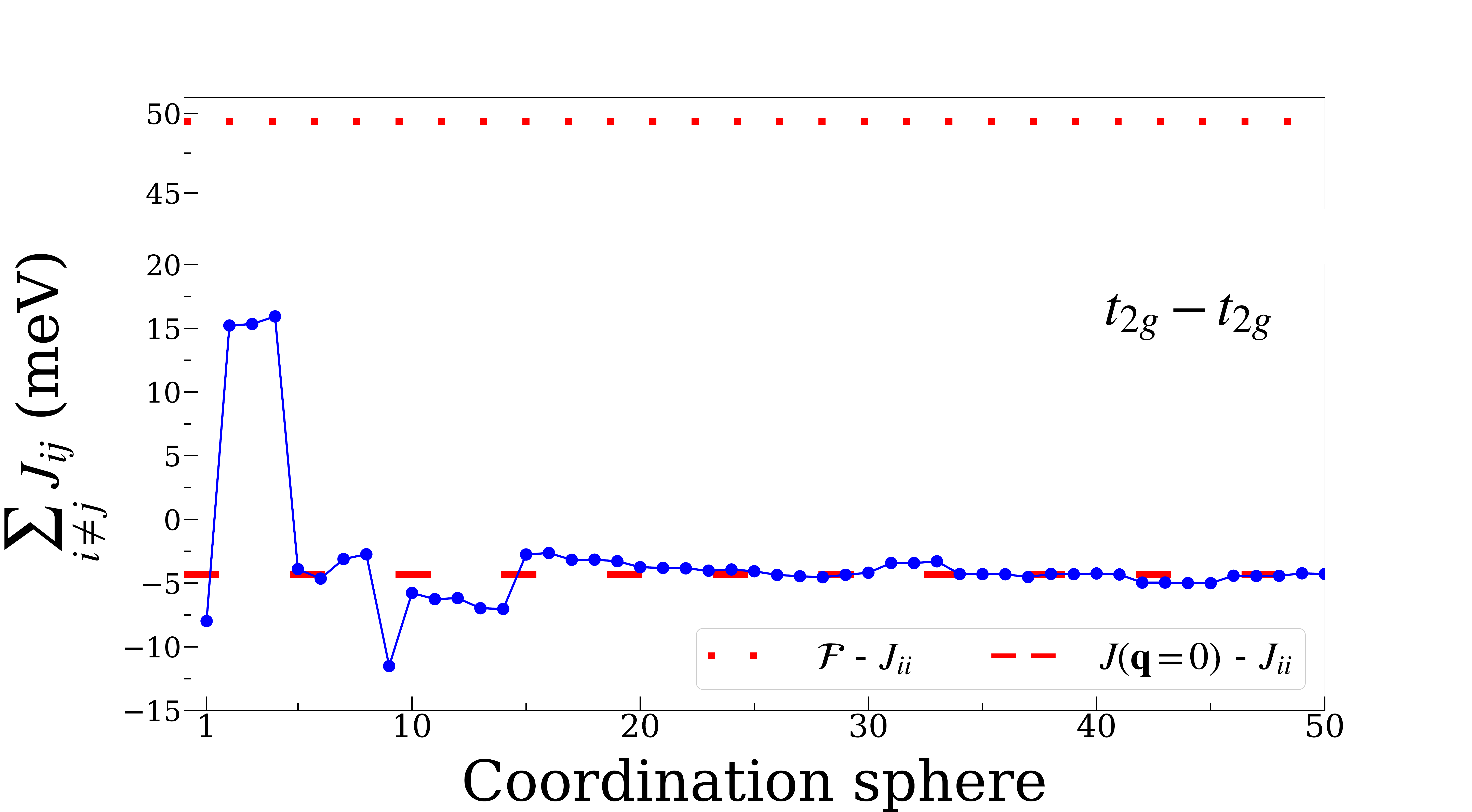}

(d)
\end{minipage}

\begin{minipage}[b]{0.49\linewidth}
\centering
\includegraphics[width = 0.99 \textwidth]{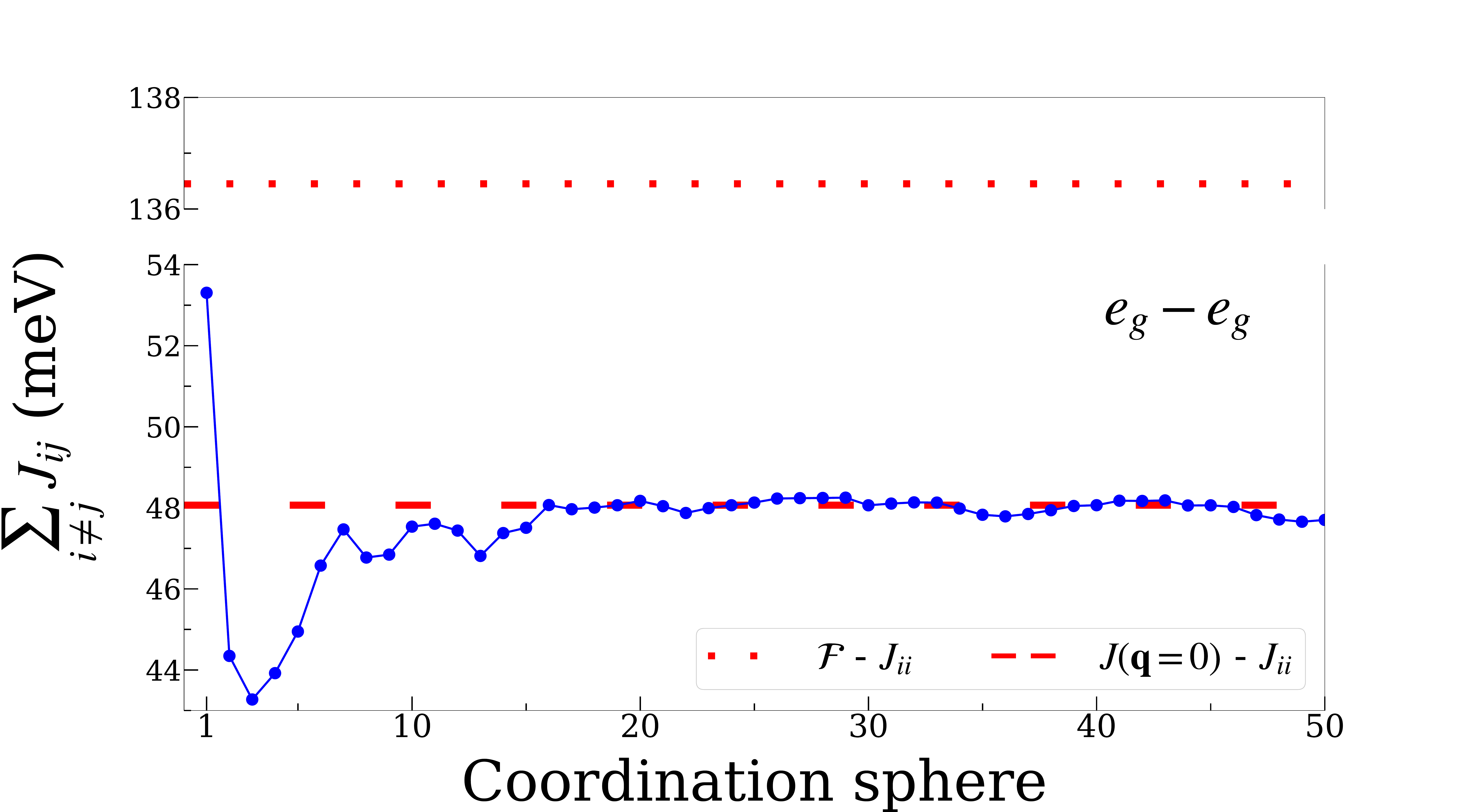}

(b)
\end{minipage}
\begin{minipage}[b]{0.49\linewidth}
\centering
\includegraphics[width = 0.99 \textwidth]{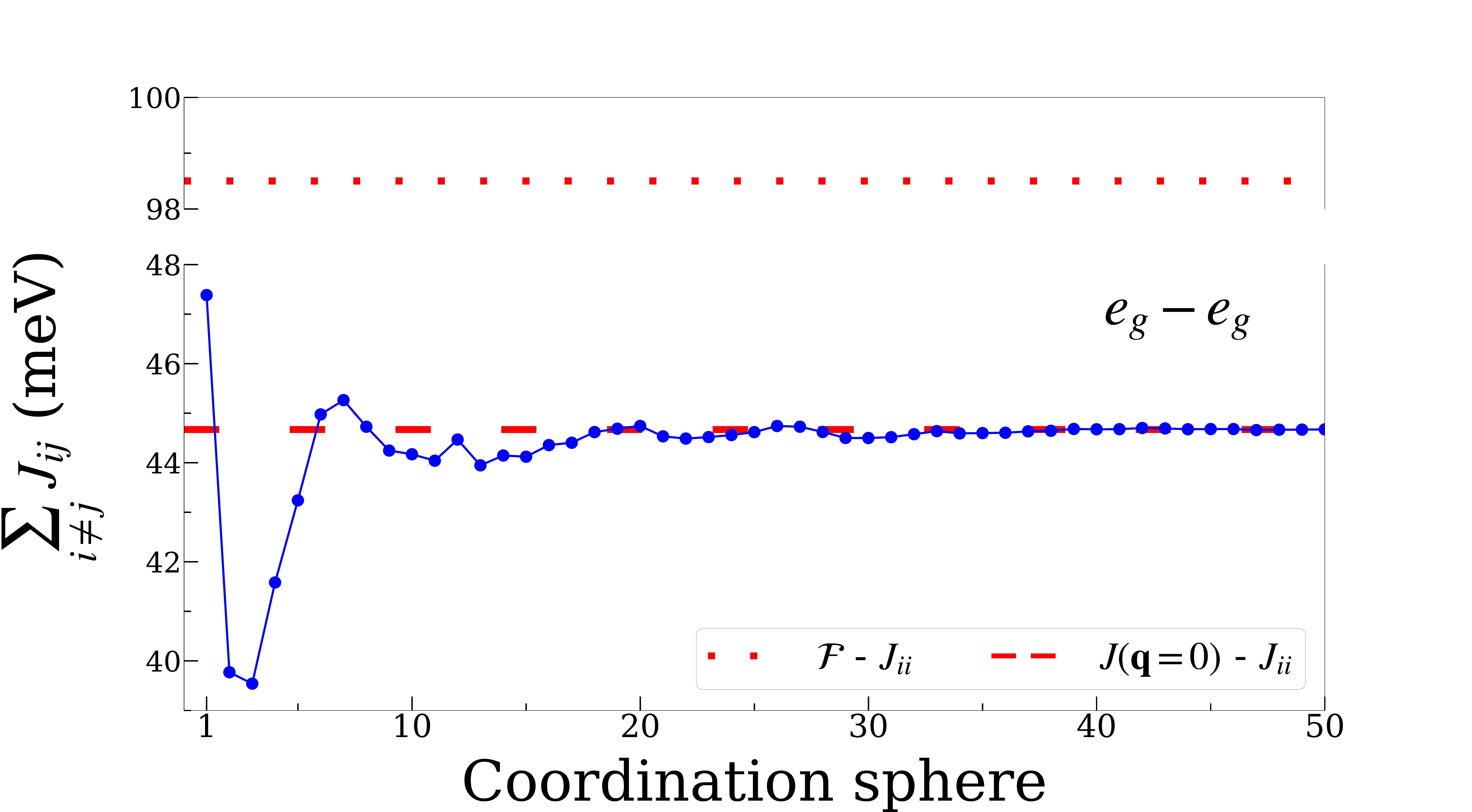}

(e)
\end{minipage}

\begin{minipage}[b]{0.49\linewidth}
\centering
\includegraphics[width = 0.99 \textwidth]{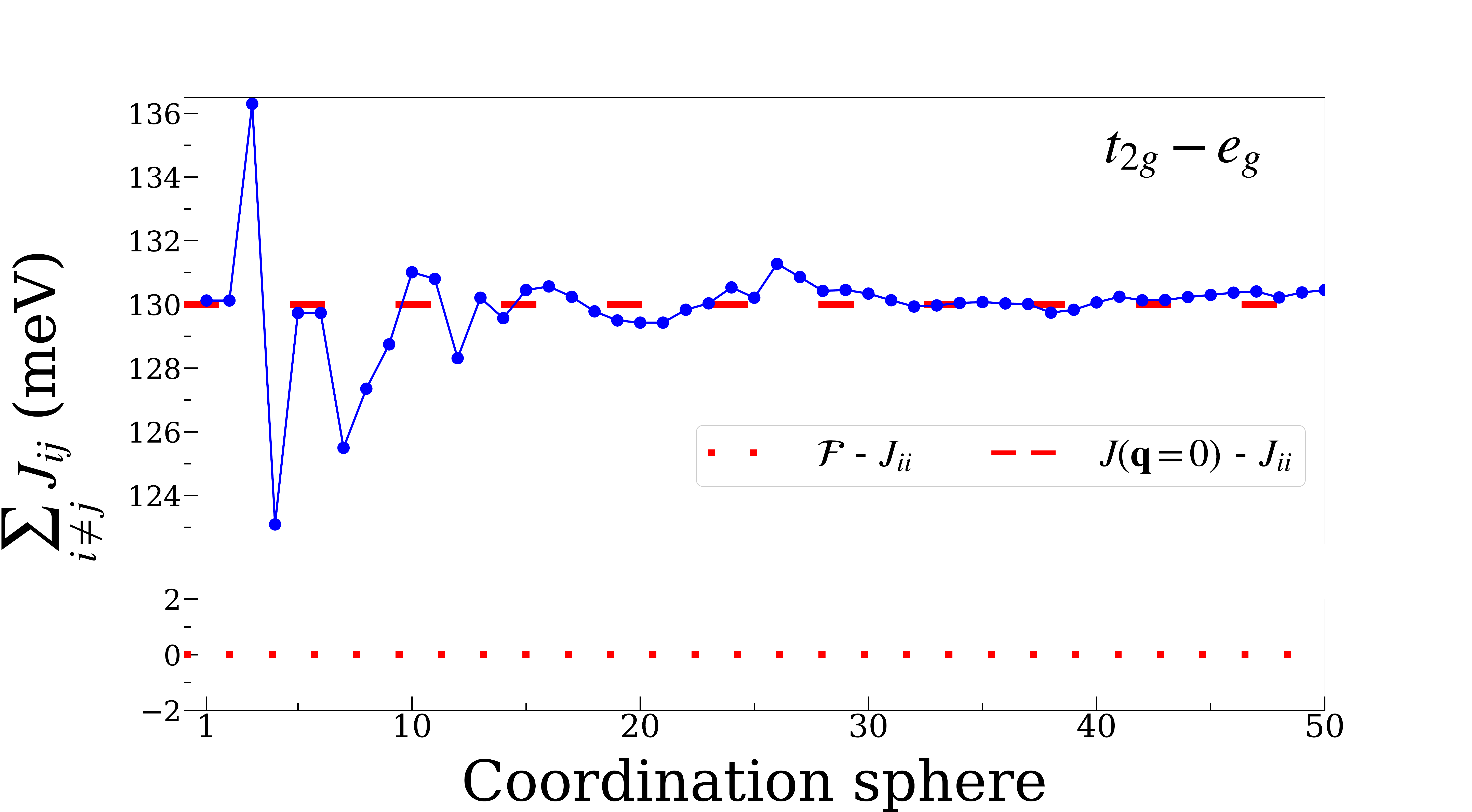}

(c)
\end{minipage}
\begin{minipage}[b]{0.49\linewidth}
\centering
\includegraphics[width = 0.99 \textwidth]{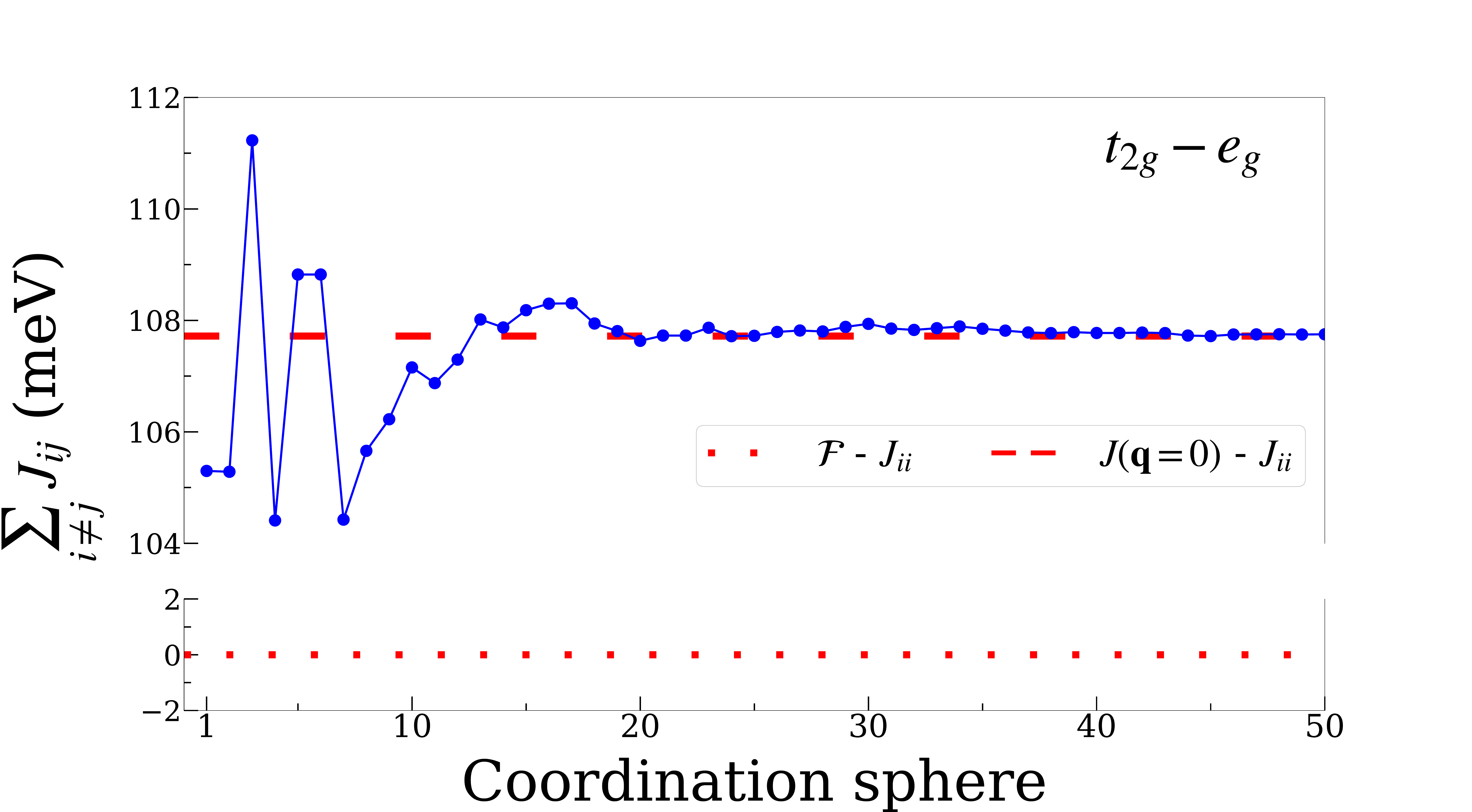}

(f)
\end{minipage}
\caption{LSDA-originated convergence dynamics of \bccFe 
(a) $\sum\limits_{j \neq i} \{ J_{ij} \}^{\ttwogttwog}$, 
(b) $\sum\limits_{j \neq i} \{ J_{ij} \}^{\egeg}$, and
(c) $\sum\limits_{j \neq i} \{ J_{ij} \}^{\ttwogeg}$ 
versus 
LDA+DMFT-originated convergence dynamics of \bccFe 
(d) $\sum\limits_{j \neq i} \{ J_{ij} \}^{\ttwogttwog}$, 
(e) $\sum\limits_{j \neq i} \{ J_{ij} \}^{\egeg}$, and
(f) $\sum\limits_{j \neq i} \{ J_{ij} \}^{\ttwogeg}$,
as more distant neighbors are taken into account.}
\label{fig:bccFe_DecomposedJij_Convergence}
\end{figure}

Let us practically study this question on the case of \bccFe. Figure \ref{fig:bccFe_DecomposedJij_Convergence} show how contributions of different symmetry orbitals to the total $J_{ij}$: 
$\ttwogttwog$,
$\egeg$,
and
$\ttwogeg$, 
individually approach the expected values, in a full accordance with the description above.
Important to note, that here assessment of these expected values can be carried out only on the level of corresponding $\JqZero - J_{ii}$, since only it inherits the orbital structure of $J_{ij}$. In Appendix we also give a full decomposition matrix $\{ J_{ij} \}^{\alpha \beta}$ for one atom couple as nearest and next-nearest neighbors, being in good agreement with the work \cite{Cardias_2017}.
It is seen that both LSDA and LDA+DMFT do not solve the problem of nonzero $\sum\limits_{j \neq i} \{ J_{ij} \}^{\ttwogeg}$, anticipated by \bccFe symmetry.


To reveal the fundamental reasons for this discrepancy, let us pay closer attention to the fact that $J(\bm{q})$ and $J_{ij}$ are related by the Fourier transform. 
Hence we can write down Parseval's identity~\cite{stade2011fourier}:
\begin{equation}
    \sum_{j}
    | J_{ij} |^2  = 
    \frac{1}{N_{\bm{q}}}
    \sum_{\bm{q}} 
    | J(\bm{q}) |^2  
    \, .
\end{equation}
By simple recombination of the terms one can obtain:
\begin{equation}
        \sum_{j \neq i}
        | J_{ij} |^2 
        - 
        \frac{1}{N_{\bm{q}}} 
        \sum_{\bm{q} \neq 0} 
        | J(\bm{q}) |^2 
        = 
        \frac{1}{N_{\bm{q}}} 
        | \JqZero |^2 
        - 
        | J_{ii} |^2 
        \, .
\end{equation}

The symmetry of the Wannier functions of actual crystals allows us to consider the decomposition matrix $\{ J_{ii} \}^{\alpha \beta}$ to be diagonal.
Thereafter, in Parseval's equality, written for $\ttwogeg$ contribution, we can replace 
$\{ \JqZero \}^{\ttwogeg}$ 
by
$\sum_{j \neq i} \{ J_{ij} \}^{\ttwogeg}$ and finally obtain:
\begin{equation}
\ParsP = \sqrt { {N_{\bm{q}}} \cdot \ParsR - \ParsQ } \, ,
\label{Basic_Parseval_Thesis}
\end{equation}
where
\begin{equation}
      \ParsP =
      \bigg|
      \sum_{j \neq i}
      \{ J_{ij} \}^{\ttwogeg}
      \bigg|
      \; ,
      \label{Pars_P} 
\end{equation}
\begin{equation}
      \ParsR =
      \sum_{j \neq i}
      \big| \{ J_{ij} \}^{\ttwogeg} \big|^2
      \; ,
      \label{Pars_R} 
\end{equation}
\begin{equation}
      \ParsQ =
      \sum_{\bm{q} \neq 0}
      \big| \{ J(\bm{q}) \}^{\ttwogeg} \big|^2 
      \label{Pars_Q} 
      \; .
\end{equation}

\begin{figure}[!h]
\includegraphics[width=0.49\columnwidth]{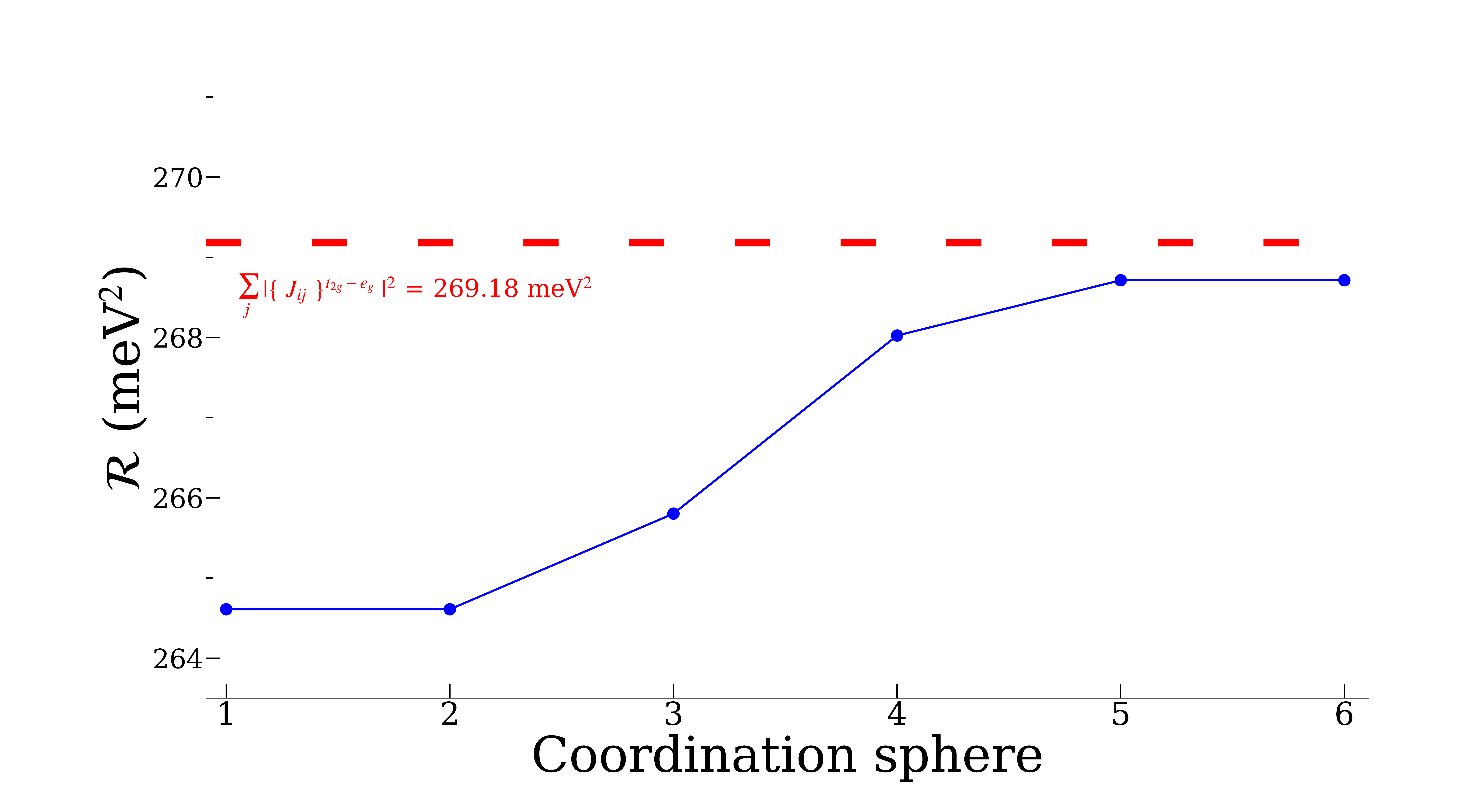}
\ 
\includegraphics[width=0.49\columnwidth]{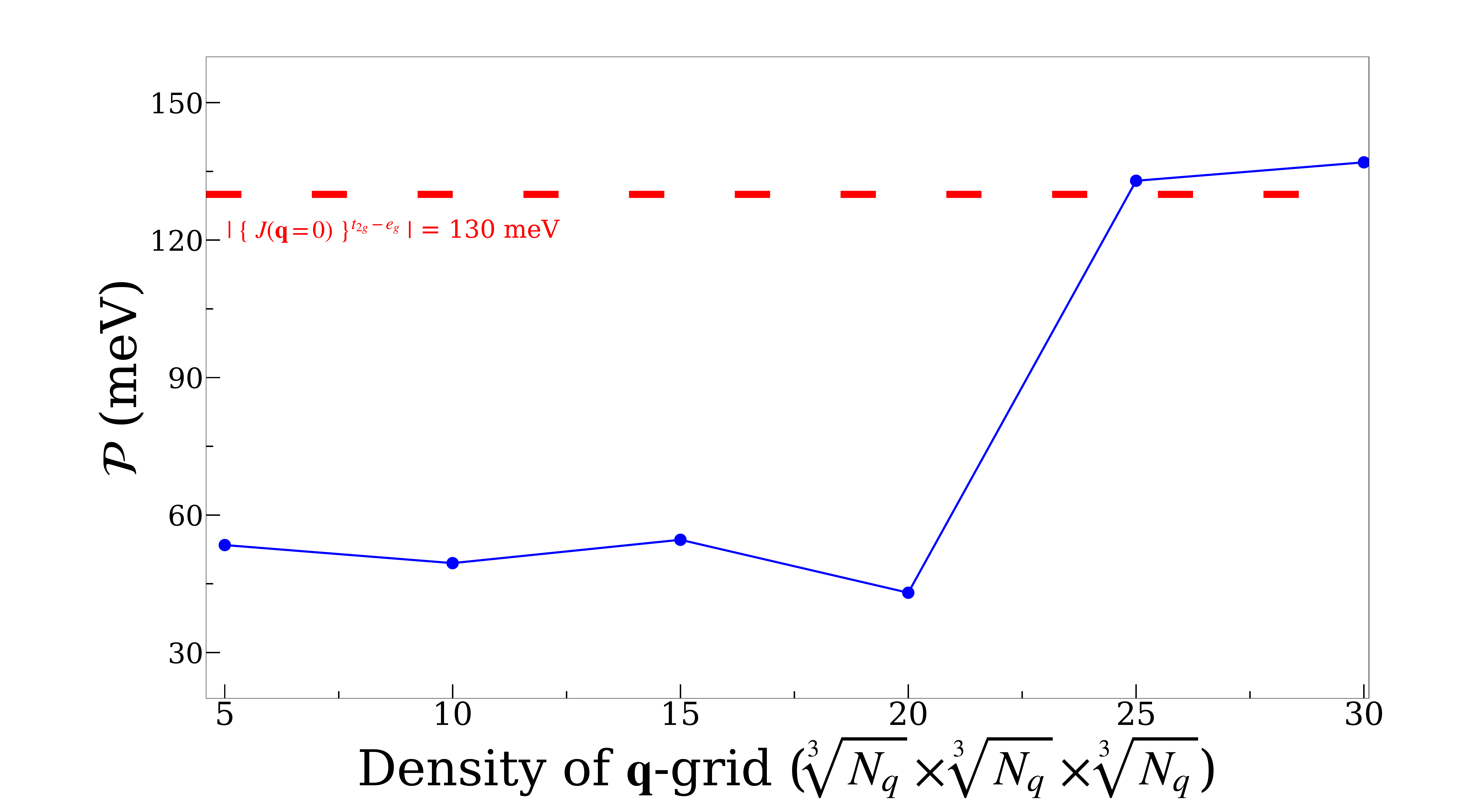}
\caption{
(\textit{Left chart}) 
$\ParsR$, Eq.~(\ref{Pars_R}), obtained for \bccFe using LSDA, as more distant neighbors are taken into account. The expected value is estimated as the combination of individual pairwise contributions up to 50\textit{th} coordination sphere.
(\textit{Right chart}) 
$\ParsP$, Eq.~(\ref{Pars_P}), obtained for \bccFe using LSDA,
as a function of $\bm{q}$-grid density 
$\QGrid$.}
\label{fig:ParsR_and_P_Convergence}
\end{figure}

Each of thus introduced parts deserves specific attention. 
It appears that $\ParsR$ has a remarkable feature:
more than 98\% of the net value (obtained as sum over 50 coordination spheres) comes from \textit{nearest neighbors}, while 6 coordination spheres cover 99.8\% (Figure \ref{fig:ParsR_and_P_Convergence} (\textit{Left chart})).

Thereafter, we can confidently state $\ParsR$ as \textit{converged constant}. In view of this fact the next step can be checking $\ParsP$ for an ability to reproduce its expected value $\big| \{ \JqZero \}^{\ttwogeg} \big|$ (130 meV), by separated means of $\bm{q}$-grid density. Figure \ref{fig:ParsR_and_P_Convergence} (\textit{Right chart}) shows that it actually takes place if the density is $25 \times 25 \times 25$ or more.
From one hand, it additionally confirms the fundamental consistency of our theoretical approach. From another hand, it alone does not lift the veil from the problem of $\ParsP$ non-suppression due to crystal's symmetry.

For the latter purpose we utilize the constancy of $\ParsR$ as the ground of the following analysis. Anticipation of $\ParsP$ = 0 in this disposition demands $\ParsQ$, Eq.~(\ref{Pars_Q}), to be \textit{linearly} proportional to $N_{\bm{q}}$. To study the actual dependency we represent $\ParsQ$ as:
\begin{equation}
      \ParsQ =
      const \cdot (N_{\bm{q}})^{\gamma}
      \, ,
\label{ParsQ_Analysis}
\end{equation}
where constant is implied to be in meV$^2$. Consequently, for $\gamma$ we got:
\begin{equation}
      \gamma = 
      \frac
    { \mathrm{d} \{ \mathrm{ln} (\ParsQ)     \} }
    { \mathrm{d} \{ \mathrm{ln} (N_{\bm{q}}) \} }
    \, .
\label{Eq_for_Gamma}
\end{equation}
Results are shown in Table \ref{Parseval_gamma}.
Here we highlight that divergence of $\gamma$ from ideal value 1 is observed on the constant level of $\sim$0.5\% (excluding the topmost point). Taking to account the coverage both areas of consistent and non-consistent $\ParsP$, we can conclude this divergence to be possessed by our theoretical formalism itself. Providing that $N_{\bm{q}} \cdot \ParsR$ and $\ParsQ$ are naturally large, it finally causes throttling the symmetry driver of $\ParsP$, at least on the computationally available grids. 
In order to affirm this statement let us approximate numerically obtained 
$\mathrm{ln} (\ParsQ) [\mathrm{ln} (N_{\bm{q}})]$ curve
by
its perfect version with $\gamma = 1$:
\begin{equation}
      \mathrm{ln} (\ParsQ)
      \approx
      \mathrm{ln} (\ParsR')
            +
      \mathrm{ln} (N_{\bm{q}})
      \, .
\end{equation}
Thus found $\ParsR' = 268.27$~meV$^2$ diverges from calculated $\ParsR$ by less than 1 meV$^2$.
It reveals the general tendency of $\ParsP$ towards suppression if we assume 
$N_{\bm{q}} \rightarrow \infty$ in extrapolating manner.

\begin{table}[h]
\begin{center}
    \begin{tabular}
    [c]{c|c|c|c}
    \hline
            $\bm{q}$-grid              &
            $\mathrm{ln} (N_{\bm{q}})$ &
            $\mathrm{ln} (\ParsQ)$     & 
            $\gamma$
        \\
    \hline
           $5 \times 5 \times 5$       &
            4.82831                    &
           10.33496                    &
            \---
        \\
          $10 \times 10 \times 10$     &
            6.90776                    &
           12.49399                    &
            1.03827
        \\
          $15 \times 15 \times 15$     & 
            8.12415                    &
           13.71624                    &
            1.00482
        \\
          $20 \times 20 \times 20$     & 
            8.98720                    &
           14.58171                    &
            1.00281
        \\
          $25 \times 25 \times 25$     & 
            9.65663                    &
            15.24779                   &
            0.99499
        \\
          $30 \times 30 \times 30$     &
           10.20359                    &
           15.79638                    &
           1.00297
        \\
    \hline
    \end{tabular}
\end{center}
\caption{Estimation of $\gamma$, Eq.~(\ref{Eq_for_Gamma}), using finite difference method.}
\label{Parseval_gamma}
\end{table}

%% file: Conclusions.tex
\section{Conclusions}

In this work we carry out long-standing problems of validated numerical modeling, if one is interested in magnetic properties of conducting materials.
It was showed analytically and numerically that the choice of magnetism's precursor has a decisive influence on the accuracy and internal consistency of the theoretical image of a real physical system.
It is important to supplement the earlier remark about the preference of using on-site sources (the Hartree-Fock method and DMFT) with a cautionary note to use \textit{combined} approaches in a manner of LSDA+DMFT for a study of characteristics, which essentially are formed as the sum of individual atomic contributions.
In addition to the naturally arising complexity of correctly accounting for double counting, the fundamental uncertainty of the source of magnetism will not allow one to control the consistency of infinitesimal spin rotations-based approaches.

We were also able to shed light on the long-known problem of symmetry breaking at the level of orbital decomposition while accumulating the spatial sums of pairwise exchange interactions in conducting systems. Consideration of the canonical \bccFe case in frame of suggested reciprocal space approach showed the origins of the vanishingly low rates of symmetry-reasoned suppression of contributions from the interaction of $t_{2g}$ and $e_g$ orbitals, leading to nonzero values of those in practical calculations.

Obtained theoretical results are believed to be of significant usefulness while elaborating the general question of long-range magnetic ordering in real metallic compounds, which stands in veil during numerical misconceptions of the present approaches.

%% file: Acknowledgements.tex
\section{Acknowledgements}
The authors acknowledge A. Szilva, I. Miranda and O. Eriksson for fruitful and inspiring discussions.
The work is supported by the grant program of the President of the Russian Federation MK-2578.2021.1.2.
The computer simulations are performed on computational resources provided by the Uran supercomputer allocated by the IMM UB RAS.

%% file: Appendix.tex
\begin{appendices}

\section{}

\subsection{The DFT-calculations, Wannier functions and tight-binding Hamiltonians}

Electronic properties of \textit{bcc} crystal structure of iron were simulated using \textit{ab initio} approach, where authors first performed LSDA calculations with exchange-correlation functional in the Perdew-Wang/Ceperley-Alder form \cite{perdew92} as implemented in the Elk code~\cite{elk-web,singh2013}. The calculation parameters are as follows. We used $a_{Fe}=2.71\,$\AA$\,$ lattice parameter and $(20 \times 20 \times 20)$ Monkhorst-Pack $\bm {k}$-point grid for the integration in reciprocal space over the Brillouin zone.

\begin{figure}[!h]
\includegraphics[width=0.49\columnwidth]{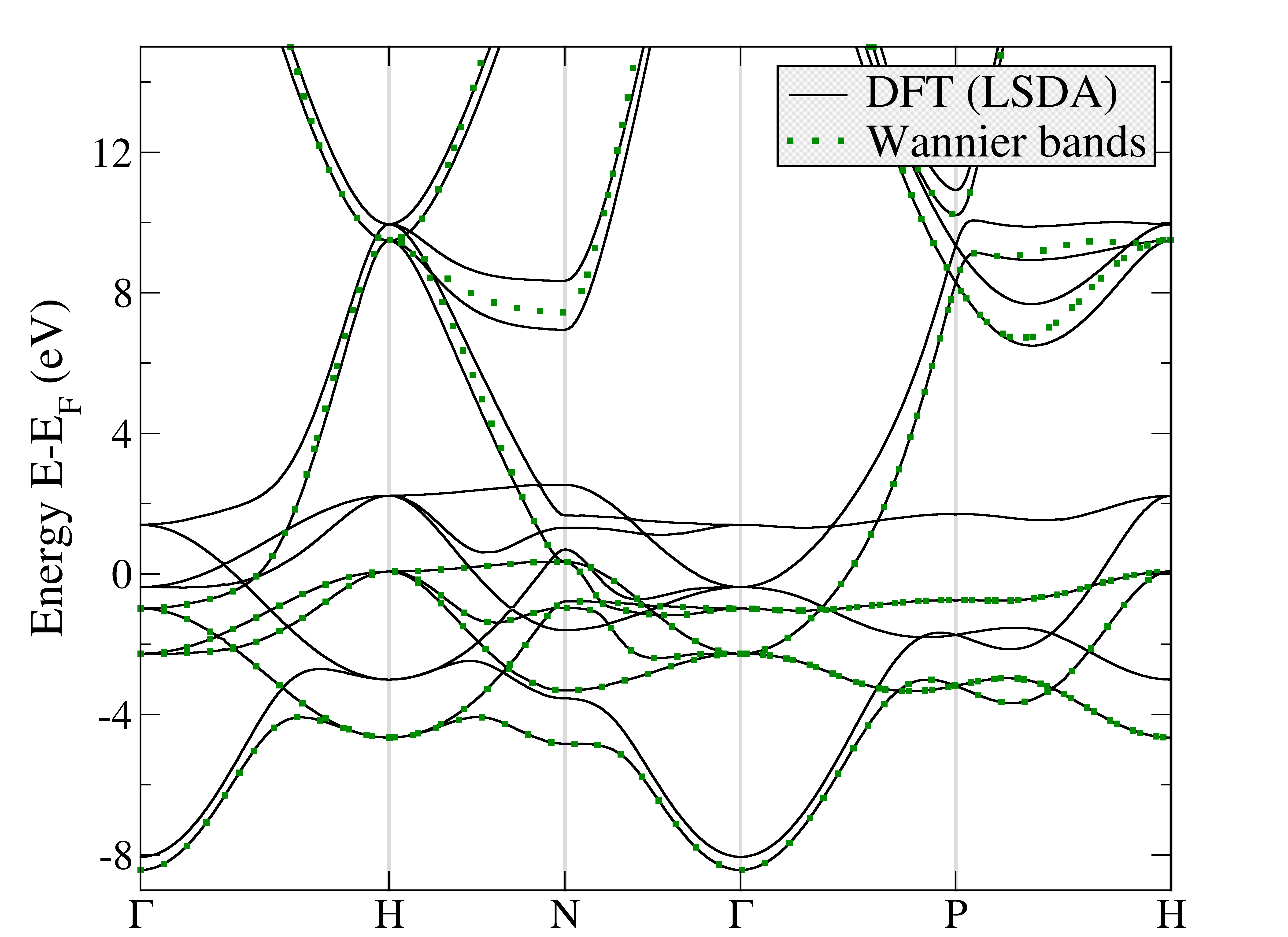}
\includegraphics[width=0.49\columnwidth]{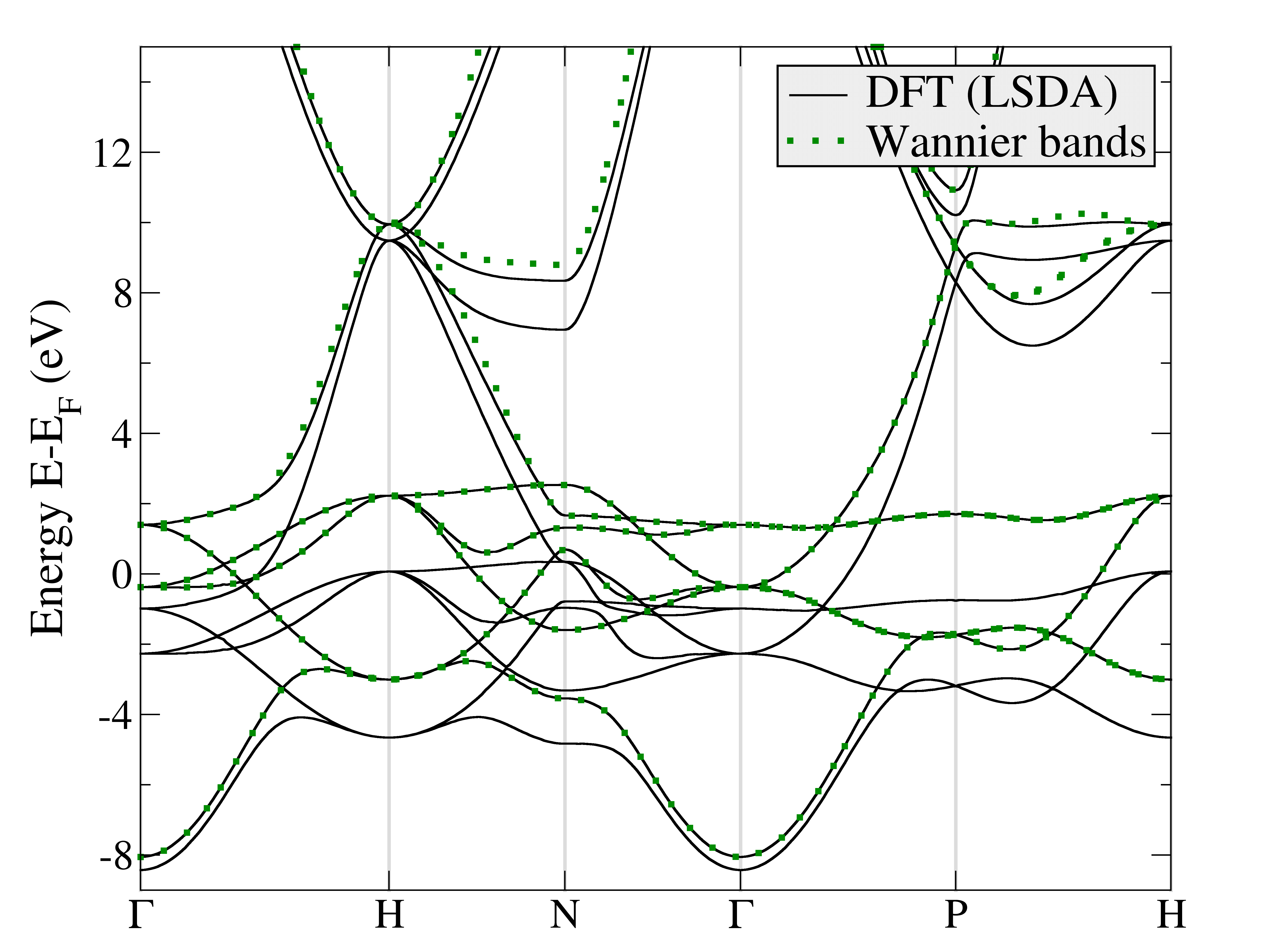}
\caption{Comparison of magnetic band structures of \textit{bcc} iron calculated using the full-potential linearised augmented-plane-waves in LSDA approach (black lines) and low-energy model in Wannier functions basis of \textit{spd} character (green dotted lines). (Left) spin-down states. (Right) spin-up states. The Fermi level corresponds to the zero energy.}
\label{fig:Fe_bands_LSDAvsWF}
\end{figure}

Figure~\ref{fig:Fe_bands_LSDAvsWF} shows the calculated spin-polarised electronic energy spectrum of the \bccFe. Afterwards, the "wannierization" procedure was applied to construct an effective Hamiltonian in the basis of the maximally localized Wannier functions~\cite{mlwf-review} for both spin states separately, where we projected the bands on the orbitals of $s$, $p$ and $d$ characters using a Wannier90 code~\cite{w90,w90-2} and the Elk to Wannier90 programming interface \cite{Gerasimov_2021}. Calculated magnetic moment per iron atom was found to be equal to 2.2 $\mu_B$, which is about experimental value~\cite{PhysRevLett.71.4067}.

Since bcc Fe with partially filled d-shell is a typical representative of intermediate correlated materials \cite{anisimov_izyumov, LDApp}, one of the most accurate approaches to describe properties of such a system is the combination of DFT and DMFT, where the electronic structure information is described by DFT while the local correlation effects are handled by DMFT \cite{vollhardt, Georges_Kotliar_1992,Georges_Kotliar_1996,kotliar_LDA_DMFT}.

In the DFT+DMFT scheme, we used DFT within the local density approximation (LDA) with non-spin-polarized \cite{perdew92} functional, as implemented in the Elk code~\cite{elk-web,singh2013}. Here we used an experimental lattice parameter for \bccFe of $a_{Fe}=2.86\,$\AA$\,$ and a $(20 \times 20 \times 20)$ Monkhorst-Pack $\bm{k}$-point mesh. To perform DMFT calculations, tight-binding Hamiltonian in the basis of maximally localized Wannier functions of \textit{spd} character on a coarser $(15 \times 15 \times 15)$ $\bm{k}$-point grid was obtained with an Elk to Wannier90 programming interface \cite{Gerasimov_2021}. 

\begin{figure}[!h]
\includegraphics[width=\columnwidth]{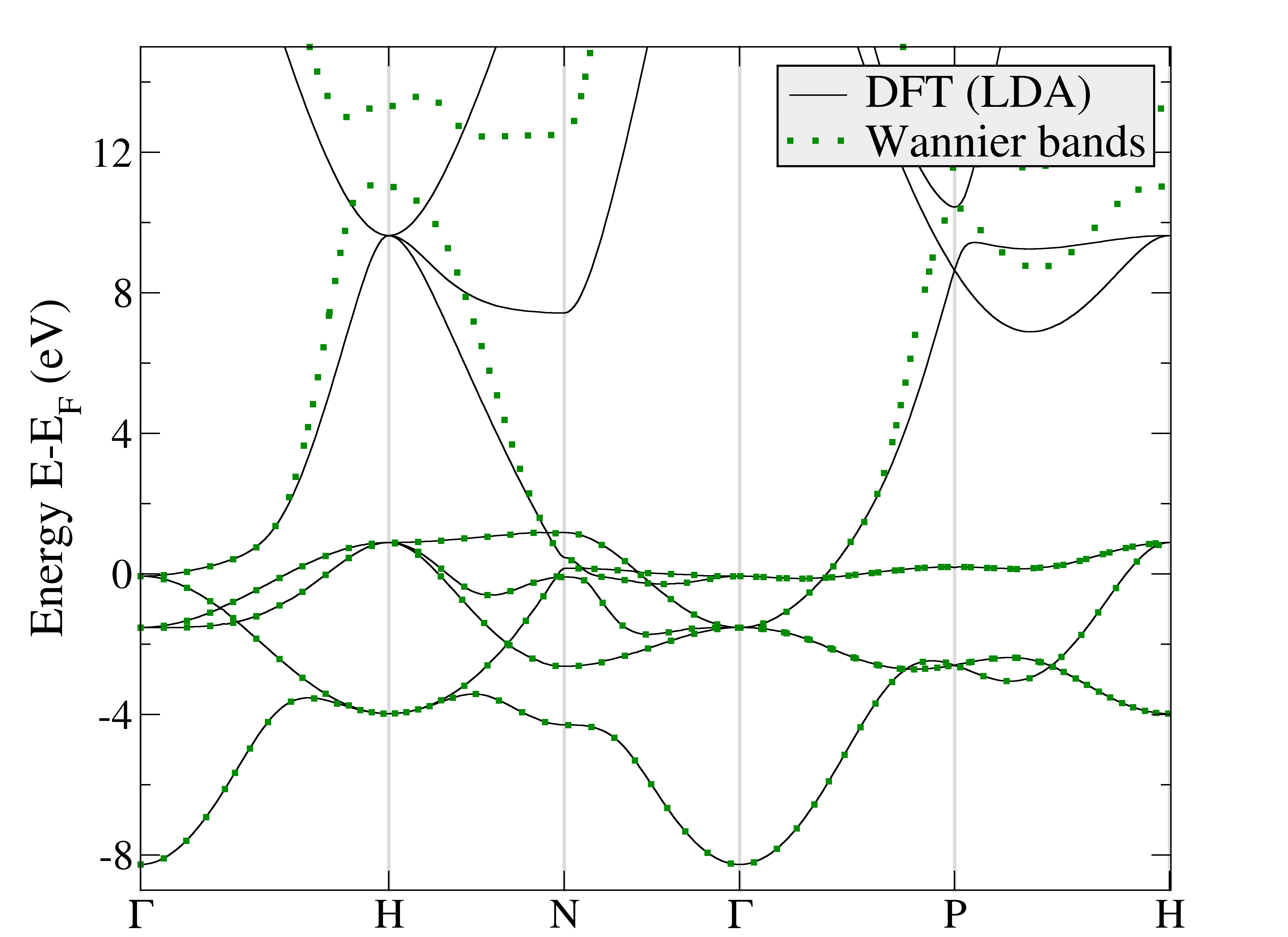}
\caption{Comparison of non-magnetic band structures of \bccFe calculated using the full-potential linearised augmented-plane-waves within LDA approximation (black lines) and low-energy model in Wannier functions basis of \textit{spd} character (green dotted lines). The Fermi level corresponds to the zero energy.}
\label{fig:Fe_bands_LDAvsWF}
\end{figure}

Figure~\ref{fig:Fe_bands_LDAvsWF} shows the LDA-derived band structure of the non-magnetic \textit{bcc} Fe. To parametrize the LDA Hamiltonian, we have constructed the low-energy model in the Wannier functions basis. Using the constructed low-energy model we have performed ferromagnetic LDA+DMFT calculations for \textit{bcc} Fe.


\subsection{DMFT calculations}

DMFT equations were solved by AMULET~\cite{amulet} toolbox. Segment version of hybridization expansion continuous-time quantum Monte Carlo (CT-QMC-HYB) solver \cite{Gull_CT-QMC-HYB} was used at $\beta$ = 10 eV$^{-1}$, where we set the Coulomb interaction parameter $U$ equal to 2.6 eV while the Hund's $J$ was set as 0.9 eV \cite{Fe_DMFT_UandJ} and appropriate double-counting (DC) correction, based on Friedel sum rule ~\cite{PhysRevB.77.205112}.

We found that the magnetic moments was stabilized at 2.17 $\mu_B$ per iron atom. Obtained value is in good agreement with previous theoretical and experimental studies~\cite{Kvashnin_2016,PhysRevLett.71.4067} as well as with the results of LSDA calculations, presented earlier in this work.

Comparing extracted using Eq.~(\ref{TwoCenteredPart}) $J_{ij}$ couplings with previous studies, we can claim a general accordance of it. Of course, one can note the deviations of obtained results within different works, at least up to the 6th coordination sphere, but this can be explained by the scheme chosen by the authors for treating magnetism in the material. The most difference happens in the 1st and the 2nd NN interactions ($J_{1NN}$ and $J_{2NN}$, respectively), where $J_{2NN}$ results are suppressed in the case of LDA+$\Sigma$ and LDA+DMFT. It is also worth to mention the 5th coordination sphere, where DMFT gains more amplitude than the others. But, as there only 8 NN on it, a significant change of the final sum of $J_{ij}$'s is not expected.

\begin{figure}[!h]
\includegraphics[width=0.99\columnwidth]{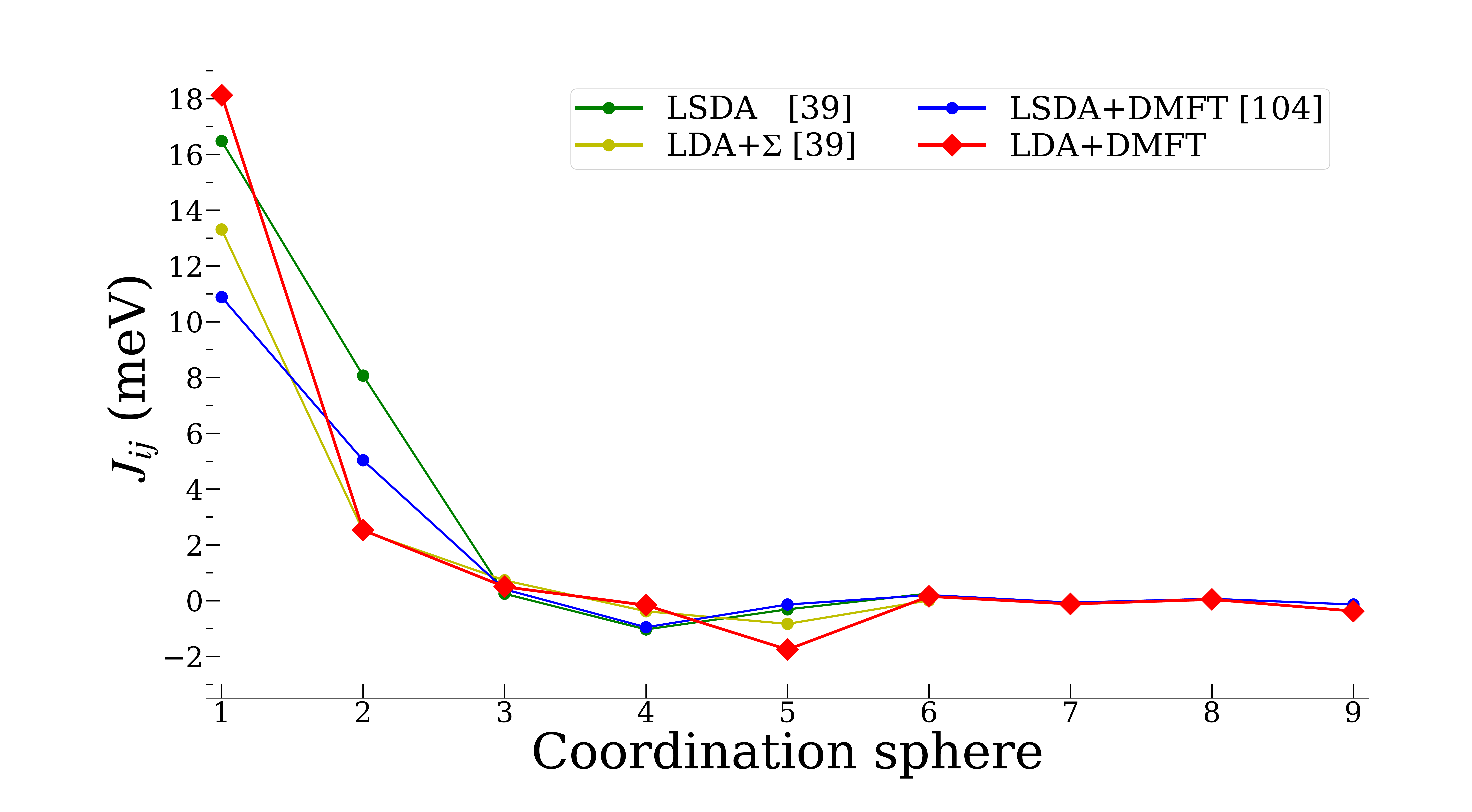}
\caption{Inter-site exchange parameters in \textit{bcc} Fe, extracted from the LDA+DMFT scheme in comparison with previous studies~\cite{LKAG_Katsnelson_2000,Kvashnin_2015_2}.}
\label{fig:Jijs_DMFT_comparison}
\end{figure}


\subsection{Orbital decomposition of \textit{bcc} Fe $J_{ij}$ for NN and Next-NN}

Investigating $J_{1NN}$ from  Table~\ref{table_J01_LSDA},~\ref{table_J01_DMFT} and \ref{table_J01_Cardias} as well as $J_{2NN}$ from Table~\ref{table_J02_LSDA},~\ref{table_J02_DMFT} and \ref{table_J02_Cardias}, we found a good agreement of obtained results with previous studies of decomposed $J_{ij}$ values of \textit{bcc} Fe.

\begin{table}[h]
\begin{center}
    \begin{tabular}
    [c]{c|c|c|c|c|c}
    \hline
                                     &
            $d_{xy}$                 &
            $d_{yz}$                 & 
            $d_{xz}$                 &
            $d_{x^2-y^2}$            &
            $d_{3z^2-r^2}$             
        \\
    \hline
            $d_{xy}$                 &
           -0.371                    &
           -1.500                    &
           -1.510                    &
            0.000                    &
            2.721                           
        \\
            $d_{yz}$                 &
           -1.495                    &
           -0.370                    &
           -1.499                    &
            2.031                    &
            0.675 
        \\
            $d_{xz}$                 &
           -1.503                    &
           -1.503                    &
           -0.367                    &
            2.031                    &
            0.672 
        \\
            $d_{x^2-y^2}$            &
            0.000                    &
            2.037                    &
            2.025                    &
            3.327                    &
            0.000
        \\
            $d_{3z^2-r^2}$           &
            2.710                    &
            0.684                    &
            0.681                    &
            0.000                    &
            3.329
        \\
    \hline
    \end{tabular}
\end{center}
\caption{Orbital-decomposed $J_{1NN}$ (in meV) in \textit{bcc} Fe obtained with LSDA, corresponding to the vector ${\bm{R}}_{ij} = ( \frac{1}{2}, \frac{1}{2}, \frac{1}{2} ) \, a$.}
\label{table_J01_LSDA}
\end{table}

\begin{table}[h]
\begin{center}
    \begin{tabular}
    [c]{c|c|c|c|c|c}
    \hline
                                     &
            $d_{xy}$                 &
            $d_{yz}$                 & 
            $d_{xz}$                 &
            $d_{x^2-y^2}$            &
            $d_{3z^2-r^2}$             
        \\
    \hline
            $d_{xy}$                 &
            0.072                    &
           -0.254                    &
           -0.243                    &
            0.000                    &
            2.254                           
        \\
            $d_{yz}$                 &
           -0.214                    &
            0.067                    &
           -0.236                    &
            1.682                    &
            0.494
        \\
            $d_{xz}$                 &
           -0.170                    &
           -0.196                    &
            0.085                    &
            1.606                    &
            0.494
        \\
            $d_{x^2-y^2}$            &
            0.000                    &
            1.617                    &
            1.689                    &
            2.962                    &
            0.000
        \\
            $d_{3z^2-r^2}$           &
            2.129                    &
            0.606                    &
            0.601                    &
            0.000                    &
            2.959
        \\
    \hline
    \end{tabular}
\end{center}
\caption{Orbital-decomposed $J_{1NN}$ (in meV) in \textit{bcc} Fe obtained with LDA+DMFT, corresponding to the vector ${\bm{R}}_{ij} = ( \frac{1}{2}, \frac{1}{2}, \frac{1}{2} ) \, a$.}
\label{table_J01_DMFT}
\end{table}

\begin{table}[h]
\begin{center}
    \begin{tabular}
    [c]{c|c|c|c|c|c}
    \hline
                                     &
            $d_{xy}$                 &
            $d_{yz}$                 & 
            $d_{xz}$                 &
            $d_{x^2-y^2}$            &
            $d_{3z^2-r^2}$             
        \\
    \hline
            $d_{xy}$                 &
           -1.333                    &
           -1.659                    &
           -1.659                    &
            0.000                    &
            2.925                           
        \\
            $d_{yz}$                 &
           -1.659                    &
           -1.333                    &
           -1.659                    &
            2.190                    &
            0.734
        \\
            $d_{xz}$                 &
           -1.659                    &
           -1.659                    &
           -1.333                    &
            2.190                    &
            0.734
        \\
            $d_{x^2-y^2}$            &
            0.000                    &
            2.190                    &
            2.190                    &
            3.836                    &
            0.000
        \\
            $d_{3z^2-r^2}$           &
            2.925                    &
            0.734                    &
            0.734                    &
            0.000                    &
            3.836
        \\
    \hline
    \end{tabular}
\end{center}
\caption{Orbital-decomposed $J_{1NN}$ (in meV) in \textit{bcc} Fe obtained in ~\cite{Cardias_2017}, corresponding to the vector ${\bm{R}}_{ij} = ( \frac{1}{2}, \frac{1}{2}, \frac{1}{2} ) \, a$.}
\label{table_J01_Cardias}
\end{table}

\begin{table}[h]
\begin{center}
    \begin{tabular}
    [c]{c|c|c|c|c|c}
    \hline
                                     &
            $d_{xy}$                 &
            $d_{yz}$                 & 
            $d_{xz}$                 &
            $d_{x^2-y^2}$            &
            $d_{3z^2-r^2}$             
        \\
    \hline
            $d_{xy}$                 &
            0.555                    &
            0.000                    &
            0.000                    &
            0.000                    &
            0.000                           
        \\
            $d_{yz}$                 &
            0.000                    &
            4.982                    &
            0.000                    &
            0.000                    &
            0.000 
        \\
            $d_{xz}$                 &
            0.000                    &
            0.000                    &
            4.960                    &
            0.000                    &
            0.000 
        \\
            $d_{x^2-y^2}$            &
            0.000                    &
            0.000                    &
            0.000                    &
            0.340                    &
            0.000
        \\
            $d_{3z^2-r^2}$           &
            0.000                    &
            0.000                    &
            0.000                    &
            0.000                    &
           -1.835
        \\
    \hline
    \end{tabular}
\end{center}
\caption{Orbital-decomposed $J_{2NN}$ (in meV) in \textit{bcc} Fe obtained with LSDA, corresponding to the vector ${\bm{R}}_{ij} = ( 0, 0, 1) \, a$.}
\label{table_J02_LSDA}
\end{table}

\begin{table}[h]
\begin{center}
    \begin{tabular}
    [c]{c|c|c|c|c|c}
    \hline
                                     &
            $d_{xy}$                 &
            $d_{yz}$                 & 
            $d_{xz}$                 &
            $d_{x^2-y^2}$            &
            $d_{3z^2-r^2}$             
        \\
    \hline
            $d_{xy}$                 &
           -0.576                    &
            0.000                    &
            0.000                    &
            0.000                    &
            0.000                           
        \\
            $d_{yz}$                 &
            0.000                    &
            2.238                    &
            0.000                    &
            0.000                    &
            0.000
        \\
            $d_{xz}$                 &
            0.000                    &
            0.000                    &
            2.216                    &
            0.000                    &
           -0.001
        \\
            $d_{x^2-y^2}$            &
            0.000                    &
            0.000                    &
            0.000                    &
            0.037                    &
            0.000
        \\
            $d_{3z^2-r^2}$           &
            0.000                    &
            0.000                    &
           -0.001                    &
            0.000                    &
           -1.217
        \\
    \hline
    \end{tabular}
\end{center}
\caption{Orbital-decomposed $J_{2NN}$ (in meV) in \textit{bcc} Fe obtained with LDA+DMFT, corresponding to the vector ${\bm{R}}_{ij} = ( 0, 0, 1) \, a$.}
\label{table_J02_DMFT}
\end{table}

\begin{table}[h]
\begin{center}
    \begin{tabular}
    [c]{c|c|c|c|c|c}
    \hline
                                     &
            $d_{xy}$                 &
            $d_{yz}$                 & 
            $d_{xz}$                 &
            $d_{x^2-y^2}$            &
            $d_{3z^2-r^2}$             
        \\
    \hline
            $d_{xy}$                 &
            0.217                    &
            0.000                    &
            0.000                    &
            0.000                    &
            0.000                           
        \\
            $d_{yz}$                 &
            0.000                    &
            4.517                    &
            0.000                    &
            0.000                    &
            0.000
        \\
            $d_{xz}$                 &
            0.000                    &
            0.000                    &
            4.517                    &
            0.000                    &
            0.000
        \\
            $d_{x^2-y^2}$            &
            0.000                    &
            0.000                    &
            0.000                    &
            0.244                    &
            0.000
        \\
            $d_{3z^2-r^2}$           &
            0.000                    &
            0.000                    &
            0.000                    &
            0.000                    &
           -1.006
        \\
    \hline
    \end{tabular}
\end{center}
\caption{Orbital-decomposed $J_{2NN}$ (in meV) in \textit{bcc} Fe obtained in ~\cite{Cardias_2017}, corresponding to the vector ${\bm{R}}_{ij} = ( 0, 0, 1) \, a$.}
\label{table_J02_Cardias}
\end{table}

\end{appendices}

%% file: Bibliography.tex
\bibliography{new_mybib}
\bibliographystyle{iopart-num}